\documentclass[review]{elsarticle}
\usepackage{amsmath,amssymb}
\usepackage{algorithmic}
\usepackage{graphicx}
\usepackage{multirow}
\usepackage{algorithm}
\usepackage{algorithmic}
\usepackage{arydshln} 
\usepackage{lineno,hyperref}

\journal{Journal of \LaTeX\ Templates}
\begin{document}

\begin{frontmatter}

\title{Multi-source adversarial transfer learning for ultrasound image segmentation with limited similarity}
\author[mymainaddress]{Yifu Zhang}
\ead{2000912@stu.neu.edu.cn}
\author[mymainaddress]{Hongru Li\corref{mycorrespondingauthor}}
\cortext[mycorrespondingauthor]{Correspondence to: College of Information Sciences and Engineering, Northeastern University, NO. 3-11, Wenhua Road, Heping District, Shenyang, 110819, China.}
\ead{lihongru@ise.neu.edu.cn}
\author[mymainaddress]{Tao Yang}
\author[mymainaddress]{Rui Tao}
\author[mymainaddress1]{Zhengyuan Liu}
\author[mymainaddress]{Shimeng Shi}
\author[mymainaddress2]{Jiansong Zhang}
\author[mymainaddress]{Ning Ma}
\author[mymainaddress]{Wujin Feng}
\author[mymainaddress]{Zhanhu Zhang}
\author[mymainaddress]{Xinyu Zhang}
\address[mymainaddress]{College of Information Sciences and Engineering, Northeastern University, Shenyang, 110819, China}
\address[mymainaddress1]{Schwarzman College, Tsinghua University, Beijing, 100084, China}
\address[mymainaddress2]{Henan Education Technology Equipment Management Center, Zhengzhou 450004, China}
\begin{abstract}
	Lesion segmentation of ultrasound medical images based on deep learning techniques is a widely used method for diagnosing diseases. Although there is a large amount of ultrasound image data in medical centers and other places, labeled ultrasound datasets are a scarce resource, and it is likely that no datasets are available for new tissues/organs. Transfer learning provides the possibility to solve this problem, but there are too many features in natural images that are not related to the target domain. As a source domain, redundant features that are not conducive to the task will be extracted. Migration between ultrasound images can avoid this problem, but there are few types of public datasets, and it is difficult to find sufficiently similar source domains. Compared with natural images, ultrasound images have less information, and there are fewer transferable features between different ultrasound images, which may cause negative transfer. To this end, a multi-source adversarial transfer learning network for ultrasound image segmentation is proposed. Specifically, to address the lack of annotations, the idea of adversarial transfer learning is used to adaptively extract common features between a certain pair of source and target domains, which provides the possibility to utilize unlabeled ultrasound data. To alleviate the lack of knowledge in a single source domain, multi-source transfer learning is adopted to fuse knowledge from multiple source domains. In order to ensure the effectiveness of the fusion and maximize the use of precious data, a multi-source domain independent strategy is also proposed to improve the estimation of the target domain data distribution, which further increases the learning ability of the multi-source adversarial migration learning network in multiple domains. The effectiveness of multi-source adversarial transfer learning is demonstrated through experiments on three datasets of ultrasound image datasets.
\end{abstract}	
\end{frontmatter}
\section{INTRODUCTION}
\label{sec:introduction}
With the development of artificial intelligence, medical image diagnosis plays an increasingly important role\cite{topol2019high}. Due to the advantages of high efficiency, non-invasiveness, non-radiation, and no need to inject contrast agents that are harmful to the human body, ultrasound medical image segmentation\cite{1661695} has become a representative method.
However, ultrasound images have low contrast and small signal-to-noise ratio, due to the blurred boundary between segmentation target and background and the inherent characteristics of a large number of echo disturbances, speckle noise and acoustic shadows\cite{van2019deep,ma2022transfer,roy2018speckle}, this results in the traditional segmentation method \cite{haralick1985image} being less effective.
\par
Deep learning \cite{lecun2015deep} represented by the U-Net network\cite{ronneberger2015u} has the potential to automatically perform ultrasound image analysis tasks\cite{LITJENS201760,long2020segmentation}. It has been used in the occasions such as breast ultrasound image segmentation \cite{guo2021segmentation}, coronary artery ultrasound image segmentation \cite{yang2019robust}. However, the high cost of labeling results in ultrasound image datasets generally having a small sample size, and deep networks can easily lead to overfitting\cite{li2019research} problems when learning directly from such datasets.
\par
Transfer learning provides a potential approach \cite{7404017} for the problem of insufficient annotation data, because there are always some similar parts \cite{shin2016deep} in things in nature. It has been applied in life expectancy\cite{zhang2021transfer,ding2022deep}, vehicle driving behavior recognition\cite{chen2023vehicles}, advertising fraud detection\cite{sisodia2022feature}, EEG pattern recognition\cite{wu2023improving} and other fields. Transfer learning makes it important to leverage medical resources from different sources \cite{yang2021weighted},
For example, Yap et al. \cite{yap2017automated} used transfer learning FCN-AlexNet to detect breast cancer ultrasound images, and achieved good results;
In the literature\cite{ayana2021transfer}, the transfer of natural images to ultrasound images is called cross-domain transfer, and the transfer between different medical images (such as between ultrasound images, between CT and ultrasound images) is called cross-modal transfer\cite{calisto2020breastscreening}.
The cross-modal transfer is easier to achieve better results than the cross-domain transfer because there are too many features in natural images that are not related to the target domain, which will lead to the extraction of redundant features that are not conducive to the task.
Song et al.\cite{song2022ct2us} used different imaging methods of the same organ for transfer learning. They used a large amount of labeled CT modal data to solve the problem of the lack of ultrasound training data with cross-modal transfer learning.
To mitigate the huge domain difference between ultrasound and CT, they employ CycleGAN\cite{zhang2021joint} to synthesize ultrasound images from CT data to construct a transition dataset \cite{kieselmann2021cross} for pre-training.
Additionally, radiologists often use previously acquired knowledge from ultrasound images of one tissue/organ type to enhance the interpretation of another type of image.
For example, calcifications have a similar appearance in ultrasound breast images and ultrasound tendon images, which helps radiologists with ultrasound breast experience quickly learn to interpret calcified tendon images.
Huang et al.\cite{huang2021cross} were inspired to try cross-modal transfer of ultrasound images between similar tissues and organs.
However, the efficiency of transfer strongly depends on the similarity between the source and target tasks, as their dissimilarity is a major contributor to negative transfer \cite{zhang2020overcoming}.
Moreover, due to ethical and other issues, there are few types of open-access ultrasound datasets, and it is difficult to obtain source-domain ultrasound datasets that are sufficiently similar to the target domain.
In addition, the boundary between the target to be segmented (such as pathological tissue) and the background (such as normal tissue) in medical images such as ultrasound is blurred and similar in texture. It is very difficult to accurately extract visual representations and improve semantic transferability\cite{9313202,8717627}.
Therefore, it is of great significance to extract as much knowledge as possible between ultrasound source domain and target domain with different sources and limited similarity.
\par
In addition to using different sources of labeled data, the unlabeled target domain data is also of great significance. Recently, adversarial thinking has been widely used to adaptively extract domain-invariant features, resulting in a method called adversarial transfer learning\cite{ganin2015unsupervised,WOS000391492800001}.
This method of evaluating the generality of deep features based on the output of the domain classifier can implicitly design a metric function, so it can avoid the designer's experience limitation and adaptively extract common deep features between domains.
In industry, it has been applied in life prediction\cite{li2022remaining}, SLAM positioning\cite{jin2020novel}, fault diagnosis, etc.\cite{han2021deep}; In terms of medical image segmentation,
 which has been applied to whole-heart segmentation\cite{liao2020mmtlnet}, optic disc segmentation\cite{liu2022cada}, mitochondrial segmentation\cite{peng2020unsupervised}, retinal anomaly segmentation\cite{wang2021weakly} and transmembrane liver segmentation\cite{hong2022unsupervised},etc.
However, to obtain enough key features in adversarial transfer learning, the source and target domains need to have sufficient similarity.
The literature \cite{de2021adversarial} points out that the properties of different domains can vary widely, especially in the medical field, where the data are heterogeneous,
because its source (eg, different patients, sensors, collection environment), different formats (eg, image resolution, sampling frequency), different scales (eg, color scale of images, units of physical or physiological measurement), or just have different probability distributions in different quantities.
Moreover, there are certain differences in the images of various organs of the human body.
In addition, there are currently fewer types of source domains that are publicly available for selection.
When faced with a certain target domain disease it may not even be possible to find a sufficiently similar source domain.
\par
\begin{figure}[!t]
	\centerline{\includegraphics[width=0.4\columnwidth]{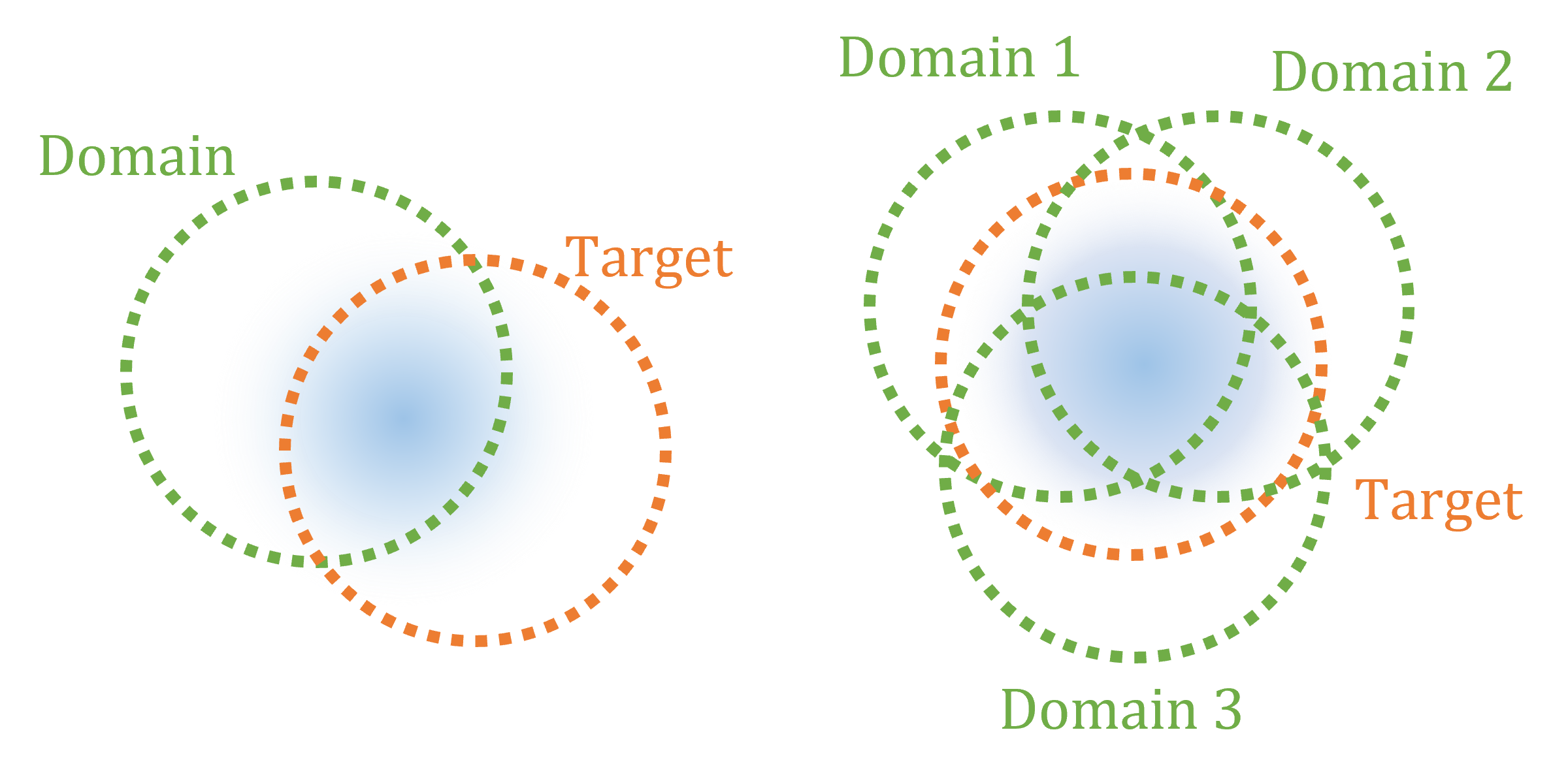}}
	\caption{A multi-source domain has the potential to provide more adequate generic features than a single source domain.}
	\label{fig_domain}
\end{figure}

Although a single source domain and target domain are not sufficiently similar, there are always similarities between different types of ultrasound images, such as contours, textures, and shadows.
Some combination of multiple source domains with different similarities to the target domain has the potential to provide sufficient knowledge \cite{mansour2008domain,yang2022multi},
Multi-source transfer learning integrates the rich knowledge of multiple source domains, which can avoid the problem of insufficient features provided by a single source domain, and reduce the impact of negative transfer\cite{gu2021integrating} as shown in Figure \ref{fig_domain}.
	One idea is to perform multi-source transfer learning \cite{bai2022three} at the data level, and some work \cite{ge2014handling,tian2022multi,zhouzhi2021multi} assigns different weights from different source domains according to the overall similarity between the source domain and the target domain. There is also work \cite{gu2021integrating} to select the most appropriate sample reweighting in multiple source domains. Another idea is to perform multi-source transfer learning at the model level. For example, multiple models are independently trained using different source domains, and the optimal model\cite{li2019multi,wu2022multi,li2022reinforcement} is selected for the task through the similarity measurement between each source domain and the target domain or according to the performance of the task. Or when facing a classification problem, vote \cite{she2022multi} for all model outputs to determine the final output. Some jobs \cite{luo2022domain,fang2021general} also design different weights for different models. Some recent works use pseudo-labels to induce domain adaptation in multiple source domains. This self-training strategy with pseudo-target labels can enhance the cross-domain capabilities of source classifiers. Methods have been developed for the data \cite{li2022dynamic} and the model \cite{dong2021confident}.
The literature \cite{zhang2021multi} has applied multi-source transfer learning to liver cancer ultrasound images.
To sum up, if natural images are used as the source domain, it is possible to extract features that are not related to the target domain, and other tissues/organs can be used for cross-modal migration. However, the ultrasound image itself has less information, and the transferable information that can be extracted is limited. Multiple source domains of ultrasound images with similar local features can be used for multi-source transfer learning to complement the lack of key features. Although ultrasound image datasets generally have the characteristics of small labeled samples, there are a large number of unlabeled ultrasound images available, so adversarial transfer learning can be used to utilize these large batches of unlabeled data. Multi-source adversarial transfer learning combines the advantages of adversarial transfer learning and multi-source transfer learning. Compared with other multi-source transfer learning, it adaptively extracts common features in multiple source domains through adversarial thinking, while providing the possibility to utilize unlabeled target domain data. Compared with adversarial transfer learning, multiple source domains avoid the problem of insufficient general features provided by a single source domain. At present, multi-source adversarial transfer learning has relatively mature solutions in classification problems, time series prediction, etc., and some people have used it in building energy prediction\cite{fang2021general}, and blood glucose prediction\cite{de2021adversarial}. There are relatively few applications in the field of medical image segmentation.
\par
To this end, a multi-source adversarial transfer learning image segmentation network for ultrasound image segmentation is proposed. Common features are extracted between multiple source domains and target domains through adversarial transfer learning to solve the insufficient target domain data labeling.
Through the multi-source fusion strategy, the problem of insufficient key features provided by a single source domain is solved.
Specifically, four components are included: feature extractor, domain classifier, feature fuser, and segmentation predictor (including source and target domains).
The feature extractor is composed of a downsampling layer, which can extract deep transferable features between domains, and provide the learned transferable features to the domain classifier and feature fusion.
The domain classifier works by distinguishing the features input by the feature extractor from the source domain or the target domain, and encourages the feature extractor to learn cross-domain representations through a gradient inversion layer.
The feature fuser fuses the information from each feature extractor and provides it to the segmentation predictor.
The segmentation predictor takes common features from the feature extractor and learns meaningful representations from the transferable features present in the source and target domains, respectively, to perform the segmentation task.
The main contributions of this paper can be summarized as follows:
\\1. To the best of our knowledge, this work is the first attempt to apply multiple source-domain guidance across tissues/organs to help solve the problem of target domain ultrasound image segmentation with insufficient label data.
\\2. We propose a multi-source adversarial transfer learning image segmentation framework and its training method, which integrates multiple adversarial transfer learning sub-networks to solve the problem that a single source domain with limited similarities cannot provide sufficient knowledge.
\\3. We design a multi-source domain-independent strategy to further enhance the network's transfer learning ability by improving the distribution estimation of the target domain and maintaining data balance while maximizing the use of source domain data.
\\The rest of this article is organized as follows. Section II details the proposed method and its background. The experimental results can be found in Section III. The conclusions are contained in Section IV.
\section{METHODS}
\subsection{Problem formulation}
Given a target domain dataset $\mathcal{D}_t^{} = \mathcal{D}_{t_l} \cup \mathcal{D}_{t_u} = \left\{ {x_j^{},y_j^{}} \right\}_{j = 1}^{n_{t_l}} \cup \left\{ {x_j^{}} \right\}_{j = n_{t_l} + 1}^{n_{t_l} + n_{t_u}}$, where, $x_j$ is the target domain image, $y_j$ is the target domain mask, $n_{t_l}$ images in $\mathcal{D}_{t_l}$ have masks, $n_{t_u}$ images in $\mathcal{D}_{t_u}$ have no masks
, the dataset $\mathcal{D}_{t}$ has a total of $n_t = n_{t_l} + n_{t_u}$ samples.
Also given a set of source domains with $N$ source domains
$\mathcal{D}_s^{} = \left\{ {\mathcal{D}_{s_i}} \right\}_{i = 1}^N$, each source domain has $n_{s_i} $ images with all pixel-level image labels, denoted as
$\mathcal{D}_{s_i} = \left\{ {x_j^i,y_j^i} \right\}_{j = 1}^{{n_{s_i}}}$, where $x_j^i $ is the source domain image, $y_j^i$ is the image mask at the pixel level of the source domain. The source domain and target domain come from different but related domains, such as ultrasound images of different organs, or images acquired by different models of ultrasound instruments.
The task considered is to transfer knowledge from $N$ ultrasound image source domains with sufficient data $\mathcal{D}_{s_i} = \left\{ {x_j^i,y_j^i} \right\}_{j = 1}^{{n_{s_i}}}$ fusion to the target domain of ultrasound images with insufficient annotation data $\mathcal{D}_t^{} = \left\{ {x_j^{},y_j^{}} \right\}_{j = 1}^{n_{t_l}} \cup \left\{ {x_j^{}} \right\}_{j = n_{t_l} + 1}^{n_{t_l} + n_{t_u}}$ to improve segmentation prediction results. Since the ultrasound images of each source domain and target domain are affected by different acquisition equipment and different source organs, their marginal probabilities present different distributions: $P\left( {{\mathcal{D}_t}} \right) \ne P\left( {{\mathcal{D}_{s_1}}} \right) \ne \cdots \ne P\left( {{\mathcal{D}_{s_i}}} \right) \ne \cdots \ne P\left( {{\mathcal{D}_{s_n}}} \right)$.
Therefore, the goal of multi-source adversarial transfer learning is to learn the transformation function $F^i_T$ by adversarial, let $P\left( F^i_T\left({{\mathcal{D}_t}}\right) \right )=P\left( F^i_T\left({{\mathcal{D}_{s_i}}}\right) \right)$, this function will be used to the segmentation predict function $\mathcal{F}_P \left({ \cdot }\right)$, in this form
$\mathcal{F}_P \left({ \cdot }\right) = F_P \left( F^i_T \left({\cdot}\right) \right)$ complete the unsupervised domain of each source domain and target domain adapt to work.
\subsection{The architecture}
\begin{figure}[!t]
	\centerline{\includegraphics[width=0.6\columnwidth]{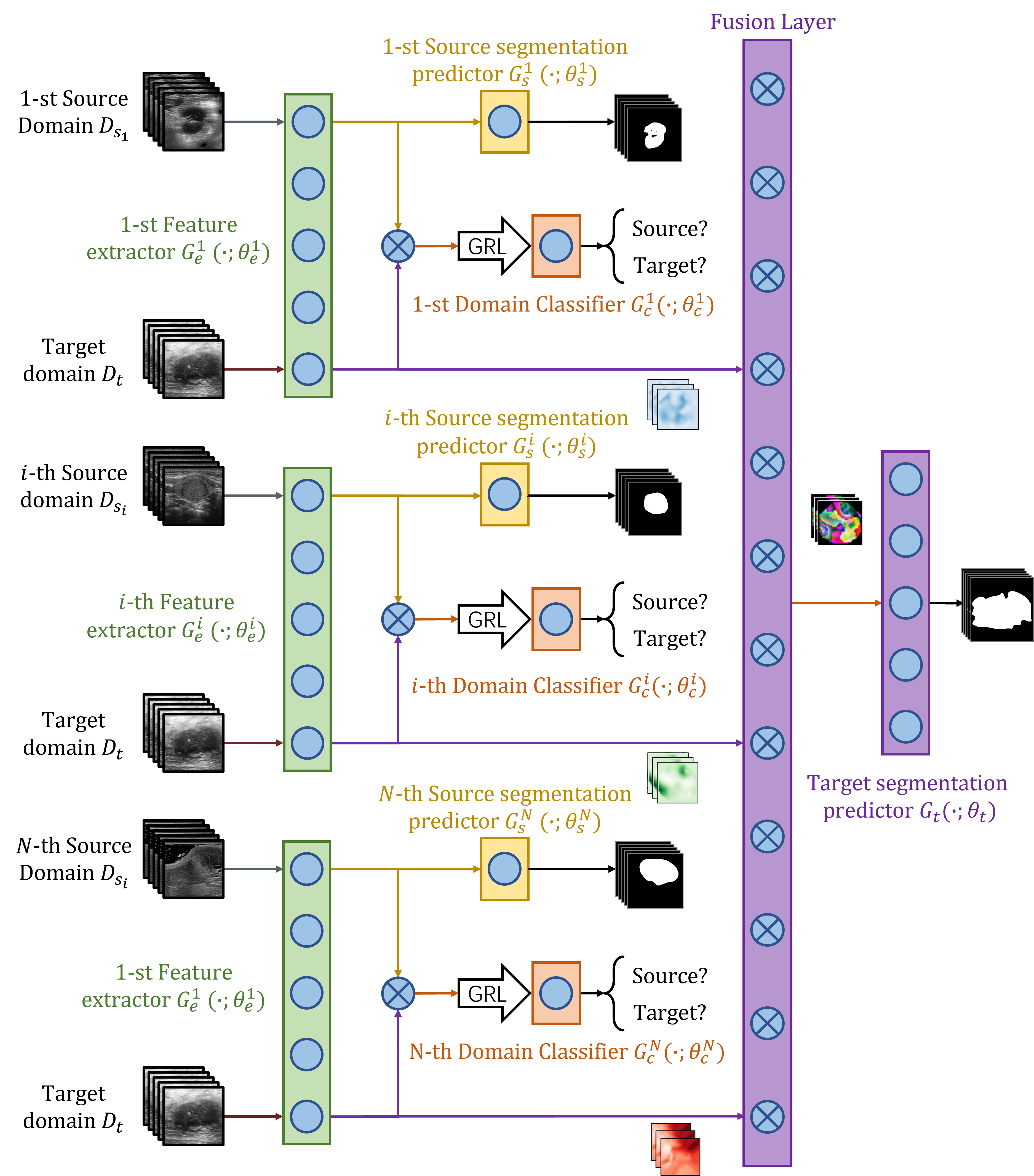}}
	\caption{Multiple Sub-networks use the feature fusion layer to form the Structure multi-source adversarial transfer learning network.}
	\label{fig_multi_source_DANN}
\end{figure}

As shown in the figure\ref{fig_multi_source_DANN}, unlike most current transfer learning methods where the source domain and the target domain share the entire model parameters, our method designs a separate sub-network for each source domain . The purpose of this is to keep the source domains independent of each other. Since the semantic information of the domains is not exactly the same, such as different types of diseases, it means that there is only partial similarity between them, and only some local features can be transferred. We hope to fuse these local features through multiple source domains to complete all the required features as much as possible. If all the parameters of the network are shared, the source domains will affect each other, so it is necessary to keep the source domains independent of each other.
The feature extractor is considered by us to be the most critical part of deep learning, because accurate extraction of available features is the basis for the task of the later task layer. Our idea is to extract the parts that are common to each source and target domain as much as possible, while keeping them as individual as possible to guarantee complementary effectiveness. Therefore, as shown in Figure \ref{fig_DANN}, in addition to a separate feature extractor, the sub-network also designs a separate source domain segmentation predictor. The parameters of each source domain and target domain are shared pairwise, and no parameters are shared between source domains at the feature extractor and source domain segmentation predictor levels. Therefore, we think that what the sub-network adaptively extracts is the local features that are common between the source domain and the target domain, rather than the features that are common between all domains.
\par
\begin{figure}[!t]
	\centerline{\includegraphics[width=0.5\columnwidth]{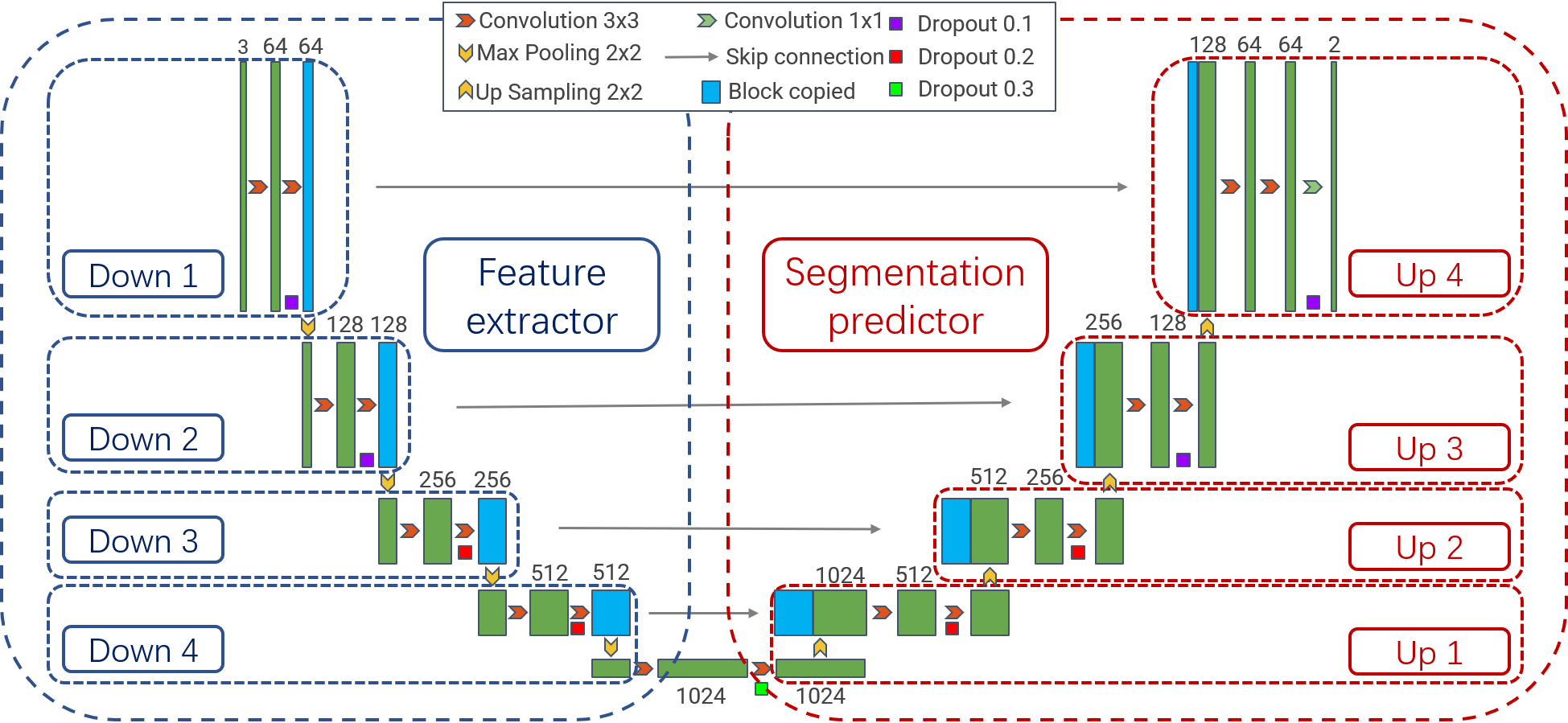}}
	\caption{U-Net has the structure of the Down-sampling Encoding, Up-sampling Decoding and Jumping Connection, which combines a more detailed graphic and deep features that help target classification.}
	\label{fig_U_Net}
\end{figure}
\begin{figure*}[!t]
	\centering
	\centerline{\includegraphics[width=0.8\columnwidth]{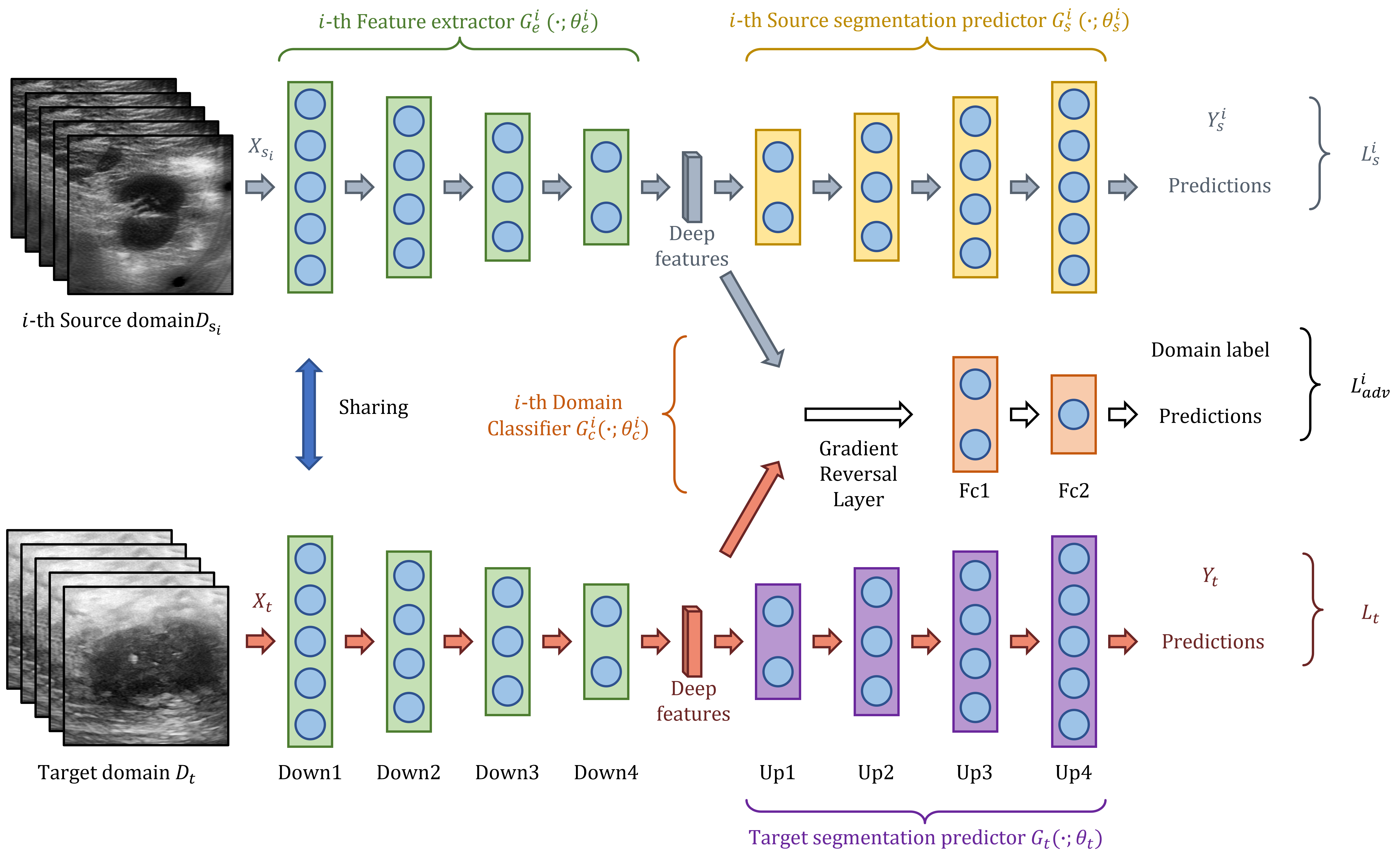}}
	\caption{The $i\mbox{-}th$ Sub-network consists of the $i\mbox{-}th$ Feature extractor $G_e^i\left( \cdot ; \theta_e^i \right)$, the $i\mbox{-}th$ Domain Classifier $G_c^i\left( \cdot ; \theta_c^i \right)$, the $i\mbox{-}th$ Source segmentation predictor $G_s^i\left( \cdot ; \theta_s^i \right)$, and the Target segmentation predictor $G_t\left( \cdot; \theta_t \right)$}
	\label{fig_DANN}
\end{figure*}

First, the sub-network of adversarial transfer learning is introduced. The main parts of the sub-network for image segmentation are the feature extractor and the segmentation predictor, which are mainly inspired by U-net, as shown in Figure \ref{fig_U_Net}, in which the encoding part performs the feature map. Down-sampling to obtain more abstract semantic features and then to judge the category of objects;
The decoding part performs Up-sampling step by step, and at the same time receives the information copied from the encoding part, and finally restores the image gradually.
\par
The adversarial transfer learning network is inspired by the generative adversarial network \cite{NIPS2014_5ca3e9b1}, and the structure is generally composed of a feature extractor, an output predictor, and a domain classifier. The common features of the source and target domains are adaptively extracted, and the common features are used as the input of the target predictor to achieve the purpose of task performance optimization in the target domain, and also provide the possibility to maximize the use of unlabeled target domain data. As shown in Figure \ref{fig_DANN}, the proposed multi-source adversarial transfer learning sub-network is divided into four parts: Feature extractor $G_e^i\left( \cdot ; \theta_e^i \right)$,
Domain classifier $G_c^i\left( \cdot ; \theta_c^i \right)$,
Target segmentation predictor $G_t\left( \cdot ; \theta_t \right)$ and Source segmentation predictor $G_s^i\left( \cdot ; \theta_s^i \right)$ where $\theta_e^i$, $ \theta_c^i$, $\theta_t$ and $\theta_s^i$ are parameters. In the vertical direction, the upper and lower parts of the Feature extractor $G_e^i\left( \cdot ; \theta_e^i \right)$ correspond to the source and target domains, respectively, and share all parameters, this part is equivalent to the same network, the image can obtain deep features after passing through the feature extractor;
The Domain classifier $G_c^i\left( \cdot ; \theta_c^i \right)$ judges whether the input comes from the source domain or the target domain according to the deep features of the input; The Source segmentation predictor $G_t\left( \cdot ; \theta_t \right)$ and the Target segmentation predictor $G_s^i\left( \cdot ; \theta_s^i \right)$ have the same structure, also accept the deep features from the feature extractor and output the predicted segmentation, but their training is done separately, as it is ensured that the tasks of the source and target domains cannot influence each other at the level of the segmentation predictor. 
Specifically, the source segmentation predictor $G_s^i\left( { \cdot ;\theta _s^i} \right)$ accepts the corresponding source domain data from this subnetwork. The target segmentation predictor ${G_t}\left( { \cdot ;{\theta _t}} \right)$ accepts data from the target domain.
Table \ref{table_net_struct} describes each predefined layer.
\begin{table}[]
	\caption{The structure definition of each layer in the network structure}   
	\label{table_net_struct}
\begin{tabular}{ccccc}
\hline
Structure               & Layer type & F/S$^a$ & P/S$^b$ & A/U$^c$      \\ \hline
\multirow{3}{*}{Down}   & MaxPool2d  & 2×2 & -   & - \\
& Conv2d     & 3×3 & 1   & ReLU     \\
& Conv2d     & 3×3 & 1   & ReLU     \\ \hdashline[1pt/1pt]
\multirow{3}{*}{Up}     & Upsample   & -   & 2   & Bilinear \\
& Conv2d     & 3×3 & 1   & ReLU     \\
& Conv2d     & 3×3 & 1   & ReLU     \\ \hdashline[1pt/1pt]
Fc & Linear     & -   & -   & ReLU     \\ \hline
\end{tabular}\\
\footnotesize{$^a$ F/S $=$ Filter/Stride\ size}\\
\footnotesize{$^b$ P/S $=$ Padding / Scale\ factor}\\
\footnotesize{$^c$ A/U $=$ Activation function / Upsampling algorithm}
\end{table}
\par
As shown in Figure \ref{fig_multi_source_DANN},
the overall network includes $N$ sub-networks, and it is necessary to comprehensively utilize the deep features of each network using  the fusion feature \cite{zhou2019review}. The purpose of this is to independently learn complementary information from different domains.
The depth features output by multiple feature extractors have chosen a fusion method similar to the add in the literature \cite{he2016deep}, 
sums abstract features from different sub-networks to make the model easier to train with less data in the target domain. The abstract features extracted by the feature extractors of multiple sub-networks will be fused in the segmentation predictor fusion layer in the target domain. Specifically, we separate out the first upsampling layer of the target domain segmentation predictor in Figure 5 to form a feature fusion layer. The fusion layer can be regarded as the transformation $g_1^{}\left( \cdot \right) = W_1^{}\left( \cdot \right)$, where $W_1^{}$ is the parameter that needs to be learned. We let the abstract features ${\cal X}_i^{}\left| {_{i = 1}^N} \right.$ from all sub-networks be summed in front of it. Then a new feature map ${\cal X}_{}^1$ is obtained through the fusion layer.
\begin{equation}
	{\cal X}_{}^1 = g_1^{}\left( {\sum\limits_{i = 1}^N {{\cal X}_i^{}} } \right) = {W_1}\left( {\sum\limits_{i = 1}^N {{\cal X}_i^{}} } \right)
	\label{fusion}
\end{equation}
The premise of being able to use this summation method is that the input data of each sub-network is exactly the same. Through the multi-source domain independent strategy designed later, we ensure that each batch of target domain data passing through each feature extractor is exactly the same. Then, after the same batch of data from the target domain passes through all sub-networks, the abstract features extracted from the same batch of data are fused in the form of summation. Then enter the second upsampling layer $g_2^{}\left( \cdot \right)$, The third upsampling layer $g_3^{}\left( \cdot \right)$, the fourth upsampling layer $g_4^{}\left( \cdot \right)$, output the predicted segmentation results. The parameters of each layer of the prediction segment $W_k^{}\left| {_{k = 1}^4} \right.$ will be updated by the marked target domain data, and all parameters will be recorded as $\theta _t$:
\begin{equation}
	\theta _t^{} = \left\{ {W_1^{};W_2^{};W_3^{};W_4^{}} \right\}
	\label{theta_t}
\end{equation}
The contribution of each source domain is implicit, because each sub-network extracts the transferable features between the corresponding source domain and the target domain during the training process through the method of adversarial transfer learning. We consider the contribution of each source domain to be determined by their similar characteristics to the target domain. 
\subsection{Multi-domain independence strategy}
Since the multi-source adversarial transfer learning includes the adversarial Domain classifier $G_c^i\left( \cdot ; \theta_c^i \right)$ , it inputs the deep features extracted by the Feature extractor $G_e^i\left( \cdot ; \theta_e^i \right)$ from the source domain $\mathcal{D}_{s_i}$ and the target domain $\mathcal{D}_t$. To ensure the adversarial effect, data balance is also very important.
Because in some cases the target domain data sample size $n_t=n_{t_l} + n_{t_u}$ is usually much smaller than the source domain data sample size $n_{s_i}$, 
if we directly mix the data of the two domains and generate batches of size $n^{b}$,
then the expected number of source domain images $E\left(n^{b}_{s_i}\right)$ is Eq.\eqref{nbsi0}, the expected number of target domain images
$E\left( n^{b}_t\right)$ is Eq.\eqref{nbt0}.
\begin{equation}
	E\left(n^{b}_{s_i}\right)=n^{b}\times{\frac{n_{s_i}}{n_t+n_{s_i}}} \label{nbsi0}
\end{equation} 
\begin{equation}
	E\left( n^{b}_t\right)=n^{b}\times{\frac{n_t}{n_t+n_{s_i}}} \label{nbt0}
\end{equation} 
Since $n_{s_i} \gg n_t$, so $E\left(n^{b}_{s_i}\right) \gg E\left( n^{b}_t\right)$. This data imbalance not only affects the learning efficiency of the target domain network, but also causes the performance of the domain classifier to degrade, which in turn affects the feature extractor's ability to extract general-purpose deep features.
\par
In addition, since all sub-networks in multi-source adversarial transfer learning are trained at the same time, it is necessary to ensure that the data from the target domain that each batch passes through each sub-network is the same, because they share the same target segmentation predictor after passing through the feature fusion layer. In most cases, the number of samples $n_{s_i}$ is different for each dataset $\mathcal{D}_{s_i}$ as the source domain. If the minimum number of source domain samples is used as the benchmark, this will lead to a large amount of valuable source domain data being wasted.
\par
\begin{figure}[!t]
	\centerline{\includegraphics[width=0.5\columnwidth]{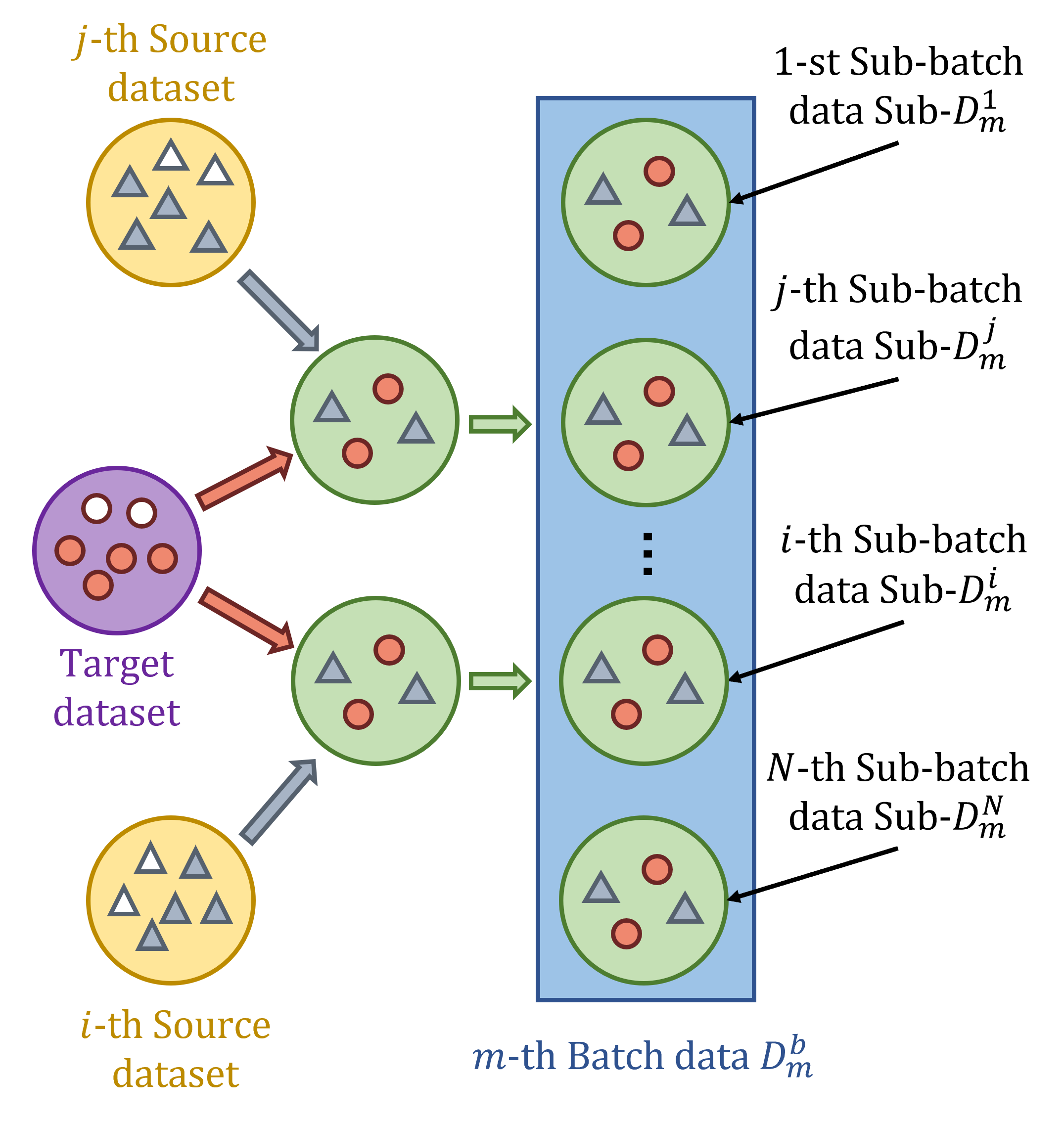}}
	\caption{Generate multiple Sub-batch data with the Multi-domain independence strategy and then constitute the batch data.}
	\label{fig_b}
\end{figure}

Domain-independent strategy \cite{YANG2021104498} is a method we proposed in the past to solve the problem of data imbalance. In addition to ensuring data balance, this method can more reliably represent the overall distribution of target domain data.
The idea of this method is to select the same number of samples in both domains to form batch data while keeping the source and target domains independent.
In this case, not only the mathematical expectation of the number of samples in the source domain $E\left( n^{b}_{s_i}\right)$ and the mathematical expectation of the number of samples in the target domain $E\left( n^{b}_t\right)$ can be the same in the batch data, but also because of the increase in the target domain data, the target domain in the batch data is also increased. The sample distribution is also closer to the overall distribution:
\begin{equation}
	E\left( n^{b}_{s_i}\right) = E\left( n^{b}_t\right) = \frac{n^{b}}{2} \label{nbsit}
\end{equation}
In this paper, to further solve the problem of waste caused by the different amounts of data in each source domain, we regard the batch data generated by the previously proposed strategy as a sub-batch data,
generate $N$ sub-batches of data for $N$ source and target domains, and then treat them as one batch.
This generalizes this strategy from the previous single source domain to multiple source domains.
At this time, the number of batch samples $n^{b}$ in Eq.\eqref{nbsit} should be rewritten as the number of sub-batch samples $n^{sb}$, and it should be guaranteed that:
\begin{equation}
	E\left( n^{sb}_{s_1}\right) = \dots = E\left( n^{sb}_{s_N}\right) = E\left( n^{sb}_t\right) = \frac{n^{sb}}{2} \label{nsbsit}
\end{equation}
Since each source domain is combined with the target domain into a sub-batch, the real batch size should be:
\begin{equation}
	n^{b} = N \times n^{sb} \label{sb}
\end{equation}
Specifically, as shown in Figure \ref{fig_b}, the same number of samples should be selected in each domain to form batch data while all source and target domains remain independent, and all data should be guaranteed to be used.
\subsection{Loss function}
In terms of the loss function of adversarial transfer learning between the target domain and the $i\mbox{-}th$ source domain $s_i$,
The target domain task is to require the target image to perform as best as possible in segmentation, and the optimization goal should be:
\begin{equation}
\mathop {\arg }\limits_{\min }\left({L_t^i}\right)=\mathop {\arg }\limits_{\min }\left({L_t^i\left( G_t\left( G_e^i\left( \cdot ; \theta_e^i \right) ; \theta_t \right) \right)}\right)
\end{equation}
Similarly, the optimization objective of the $i\mbox{-}th$ source domain $s_i$ should be:
\begin{equation}
\mathop {\arg }\limits_{\min }\left({L_s^i}\right)=\mathop {\arg }\limits_{\min }\left({L_s^i\left( G_s^i\left( G_e^i\left( \cdot ; \theta_e^i \right) ; \theta_s^i \right) \right)}\right)
\end{equation}
The purpose of adversarial is to extract general features, and the domain classifier is required to be as indistinguishable as possible from the source of the extracted features. The optimization goal should be:
\begin{equation}
\mathop {\arg }\limits_{\max }\left({L_{adv}^i}\right)=\mathop {\arg }\limits_{\max } L_{adv}^i \left( G_c^i\left( G_e^i\left( \cdot ; \theta_e^i \right) ; \theta_c^i \right) \right)
\end{equation}
 To unify the form with the segmentation task to facilitate the solution, it should be written in the form of $\mathop {\arg }\limits_{\min }{L_{adv}^i}$.
According to the literature \cite{WOS000391492800001}, a Gradient Reversal Layer (GRL) is introduced to multiply the gradient by some negative constant (such as a negative identity matrix $-I$) during backpropagation-based training. The form of adversarial transfer learning optimization objective for a single source domain versus the target domain is:
\begin{equation}
\mathop {\arg }\limits_{\min }\left({\mathcal{L}_i}\right)= \mathop {\arg }\limits_{\min }\left({L}_s^i+\alpha {L}_t^i-\lambda {L}_{adv}^i \right)
\end{equation}%
Since the target domains in the same batch may not all have labels, the above formula needs to be refined for the batch, and then the overall loss function can be obtained. First, starting from the sample level, let the sample $X_j$ come from the source domain ${D}_{s_i}$, that is, $X_j \in \mathcal{D}_{s_i}$, then the loss function for this sample is:
\begin{equation}
	\begin{aligned}
		\mathcal{L}_i\left( X_j \right) 
		&= {L}_s^i-\lambda {L}_{adv}^i \\
		&={L_s^i\left( G_s^i\left( G_e^i\left( X_j ; \theta_e^i \right) ; \theta_s^i \right) \right)} \\
		&-\lambda L_{adv}^i \left( G_c^i\left( G_e^i\left( X_j ; \theta_e^i \right) ; \theta_c^i \right) \right)
	\end{aligned}
\label{ls}
\end{equation}
Similarly, if the sample comes from the labeled target domain, that is, $X_j \in \mathcal{D}_{t_l}$, the loss function for this sample is:
\begin{equation}
	\begin{aligned}
		\mathcal{L}_i\left( X_j \right) 
		&={L}_s^i-\lambda {L}_{adv}^i \\
		&={L_s^i\left( G_t\left( G_e^i\left( X_j ; \theta_e^i \right) ; \theta_t \right) \right)} \\
		&-\lambda L_{adv}^i \left( G_c^i\left( G_e^i\left( X_j ; \theta_e^i \right) ; \theta_c^i \right) \right)
	\end{aligned}
\label{ltl}
\end{equation}
Assuming that the sample comes from the marked target domain $X_j \in \mathcal{D}_{t_u}$, the loss function for this sample is:
\begin{equation}
	\begin{aligned}
		\mathcal{L}_i\left( X_j \right) 
		&= -\lambda {L}_{adv}^i \\
		&=-\lambda L_{adv}^i \left( G_c^i\left( G_e^i\left( X_j ; \theta_e^i \right) ; \theta_c^i \right) \right)
	\end{aligned}
\label{ltu}
\end{equation}
Integrate Eq.\eqref{ls}, Eq.\eqref{ltl}, and Eq.\eqref{ltu}, the loss function for a single sub-batch $Sub\mbox{-}D_{m}^i$ containing the $i\mbox{-}th$ source domain, the labeled target domain, and the unlabeled target domain is:
\begin{equation}
\begin{aligned}
\mathcal{L}_{b}^i 
&= {\sum\nolimits_{X_j \in \left( Sub\mbox{-}D_{m}^i \cap \mathcal{D}_{s_i} \right)}L_s^i\left( G_s^i\left( G_e^i\left( X_j ; \theta_e^i \right) ; \theta_s^i \right) \right)}\\
&+\alpha{\sum\nolimits_{X_j \in \left( Sub\mbox{-}D_{m}^i \cap \mathcal{D}_{t_l} \right)}L_t\left( G_t\left( G_e^i\left( X_j ; \theta_e^i \right) ; \theta_t \right) \right)}\\
&-\lambda \sum\nolimits_{X_j \in Sub\mbox{-}D_{m}^i}L_{adv}^i \left( G_c^i\left( G_e^i\left( X_j ; \theta_e^i \right) ; \theta_c^i \right) \right)
\end{aligned}
\label{loss}
\end{equation}
Combined with \eqref{loss} generalizing the number of source domains from $1$ to $N$, the loss function for all source domains is:
	\begin{equation}
		\begin{aligned}
		\mathcal{L}&=\sum\nolimits_{i=1}^N\mathcal{L}_{b}^i \\
		&= \sum\nolimits_{i=1}^N{\sum\nolimits_{X_j \in \left( Sub\mbox{-}D_{m}^i \cap \mathcal{D}{s_i} \right)}L_s^i\left( G_s^i\left( G_e^i\left( X_j ; \theta_e^i \right) ; \theta_s^i \right) \right)}\\
		&+\alpha{\sum\nolimits_{X_j \in \left( Sub\mbox{-}D_{m}^i \cap \mathcal{D}_{t_l} \right)}L_t\left( G_t\left( \sum\nolimits_{i=1}^N G_e^i\left( X_j ; \theta_e^i \right) ; \theta_t \right) \right)}\\
		&-\lambda \sum\nolimits_{i=1}^N \sum\nolimits_{X_j \in Sub\mbox{-}D_{m}^i}L_{adv}^i \left( G_c^i\left( G_e^i\left( X_j ; \theta_e^i \right) ; \theta_c^i \right) \right)
	\end{aligned}
	\label{losstotal}
\end{equation}
\subsection{The learning process}
The purpose of the proposed multi-source adversarial transfer learning is to extract the common underlying features of each source domain $D_{s_i}$ and target domain $D_t^{}$ in the source domain set $D_s^{} = \left\{ {D_{s_i}} \right\}_{i = 1}^N$ in an adversarial manner.
The domain-independent strategy is extended to multiple source domains to ensure that each valuable source domain data is not wasted while ensuring balanced learning of each sample in each source domain and target domain.
For training this network, the idea is to train multiple Feature extractors $G_e^i\left( X_j ; \theta_e^i \right)$ to extract deep features.
For general feature extraction, data from the source domain and the target domain will be given Domain Labels, and then enter the Domain classifier $G_c^i\left( G_e^i\left( X_j ; \theta_e^i \right) ; \theta_c^i \right)$ to complete the domain adaptation.
The sub-batch data from the source domain trains the Source segmentation predictor $G_s^i\left( G_e^i\left( X_j; \theta_e^i \right); \theta_t \right)$ of the sub-network after the feature extractor and obtains the result of the source domain image segmentation.
The sub-batch data from the target domain is extracted and the Target segmentation predictor $G_t\left( \sum\limits_{i = 1}^N G_e^i\left( X_j ; \theta_e^i \right) ; \theta_t \right)$ of the sub-network is trained, and the result of the target domain image segmentation is obtained.
Therefore, the optimization objective is:
\begin{equation}
	\left\{\widehat{\theta_e^i};\widehat{\theta_t};\widehat{\theta_s^i};\widehat{\theta_c^i}\right\}_{i=1}^N = \mathop {\min }\limits_{ \Theta }\left({\mathcal{L}\left(\Theta\right)}\right)
\end{equation}
Where, $\Theta=\left\{\theta_e^i;\theta_t;\theta_s^i;\theta_c^i\right\}_{i=1}^N$ is the network parameter, $\widehat{\Theta}=\left\{\widehat{\theta_e^i};\widehat{\theta_t};\widehat{\theta_s^i};\widehat{\theta_c^i}\right\}_{i=1}^N$ is the estimated parameter of the network.
After the network is trained, the learned transformation function is $F^i_T=G_e^i\left(\cdot; \theta_e^i \right)$, which can extract common features between the source and target domains. Segmentation predict function $F_P=G_t\left( \cdot; \theta_t \right)$ utilizes the learned general features and has good generalization ability. Finally, the prediction function can be directly applied to the segmentation prediction task of the target ultrasound image.
\begin{equation}
\widehat{\mathcal{F}_P} \left( \cdot \right)=G_t\left(\sum\limits_{i = 1}^N G_e^i\left( \cdot; \theta_e^i \right); \theta_t \right)
\end{equation}
The specific implementation steps are shown in the algorithm \ref{alg:Framwork}
\begin{algorithm}[h]
	\caption{Multi-source Adversarial Transfer Learning Algorithm}
	\label{alg:Framwork}
	\begin{algorithmic}[1]	 
		\REQUIRE ~~\\ 
		The source datasets, $D_s^{} = \left\{ {D_{s_i}} \right\}_{i = 1}^N$,
		Where, $D_{s_i} = \left\{ {x_j^i,y_j^i} \right\}_{j = 1}^{{n_{s_i}}}$;\\
		The target dataset, $D_t^{}$,
		Where, $D_t^{} = \left\{ {x_j^{},y_j^{}} \right\}_{j = 1}^{n_{t_l}} \cup \left\{ {x_j^{}} \right\}_{j = n_{t_l} + 1}^{n_{t_l} + n_{t_u}}$;\\
		Total epoch, $e_{total}$;\\
		Sub-batch size, $n^{sb}$; 
		
		\STATE Multi-domain independence strategy 
		\FOR {$e=1$; $e \leq e_{total}$; $e++$}
		\FOR{$m=1$; $m \leq \max\left(n_{s_i|_{i=1}^N}\right)/n^{sb} $; $m++$}
		\FOR{$i=1$; $i \leq N$; $i++$}
		\STATE Extract deep features $G_e^i\left( X^i_m; \theta_e^i \right)$ from the batch dataset $D^i_m$
		\STATE Get predictions $G_s^i\left( G_e^i\left( X^i_m; \theta_s^i \right); \theta_d^i \right)$, $G_t\left( G_e^i\left( X^i_m; \theta_e^i \right); \theta_t \right)$ and $G_c^i\left( G_e^i\left( X^i_m; \theta_e^i \right); \theta_c^i \right)$;
		\ENDFOR
		\STATE Calculate the loss function as Eq.\eqref{losstotal};
		\STATE Backpropagation to update the network parameters $\Theta=\left\{\theta_e^i;\theta_t;\theta_s^i;\theta_c^i\right\}_{i=1}^N$ from $\mathcal{L}_{b}$;
		\ENDFOR
		\ENDFOR
		\ENSURE ~~\\ 
		Deep prediction model, $\widehat{\mathcal{F}_P} \left( \cdot \right)=G_t\left( \sum\limits_{i = 1}^N G_e^i\left( \cdot; \theta_e^i \right); \theta_t \right)$;
	\end{algorithmic}
\end{algorithm}

\section{EXPERIMENTAL RESULTS AND ANALYSIS}\label{exp}
\subsection{Dataset description}
\begin{figure}[!t]
	\centerline{\includegraphics[width=\columnwidth]{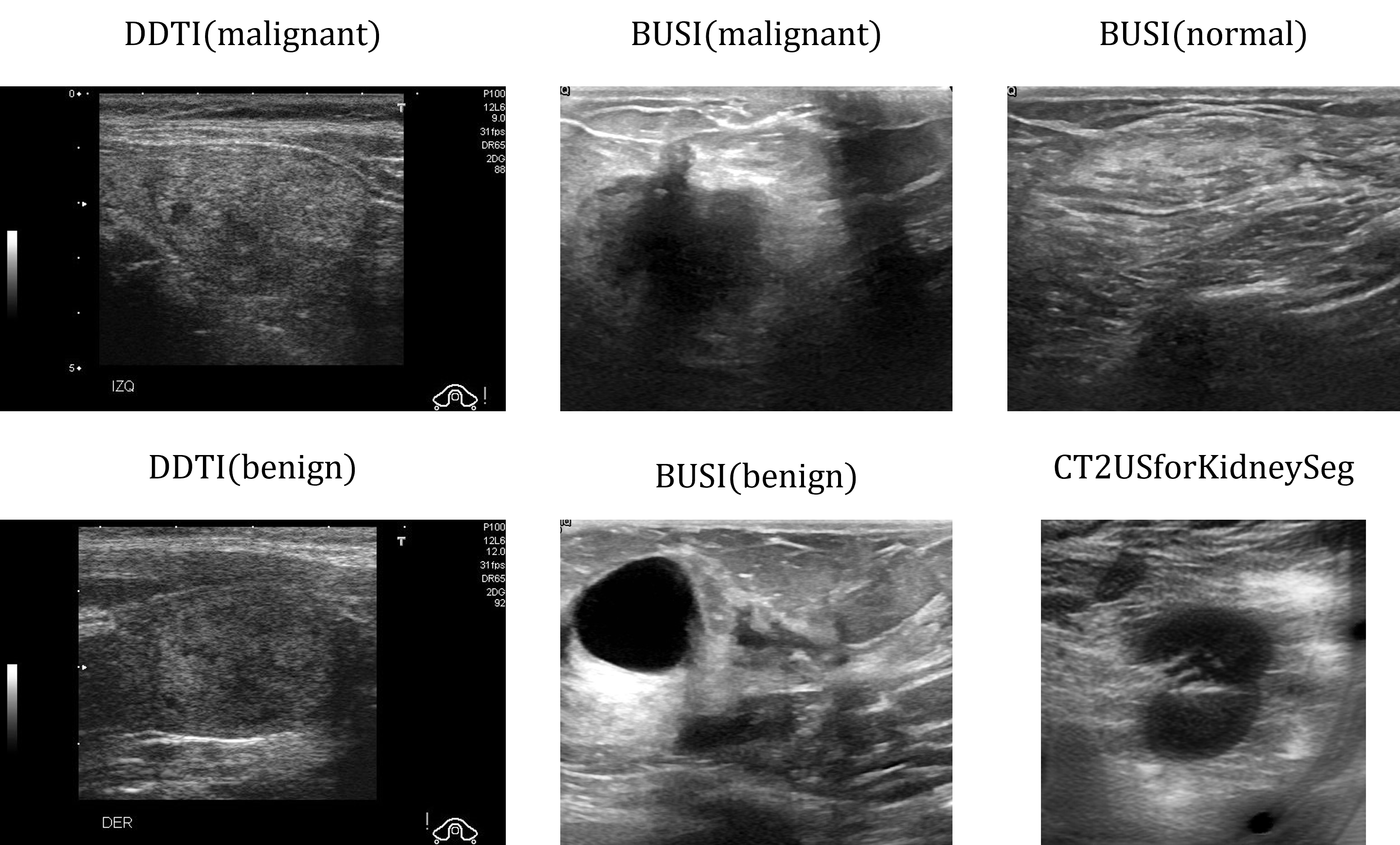}}
	\caption{Examples of 6 types of raw data in the Breast Ultrasound Images Dataset, DDTI Dataset and CT2USforKidneySeg Dataset.}
	\label{data}
\end{figure}

In this paper, ultrasound images from three datasets are used to test the performance of multi-source adversarial transfer learning in ultrasound image segmentation, involving organs including the breast\cite{al2020dataset}(Breast Ultrasound Images Dataset, Dataset BUSI), Thyroid (DDTI) \cite{pedraza2015open} and kidney (CT2USforKidneySeg) \cite{song2022ct2us}, their unprocessed raw examples are shown in Figure \ref{data}, and these datasets are described below.
\subsubsection{Breast Ultrasound Images Dataset}
The \href{https://scholar.cu.edu.eg/Dataset_BUSI.zip}{BUSI} dataset
 \cite{al2020dataset} was collected in 2018 and includes breast ultrasound images of 600 women between the ages of 25 and 75. The dataset consists of 780 images in PNG format with an average image size of 500 × 500 pixels. The images are divided into three categories: normal, benign and malignant, with 133, 437 and 210 images, respectively.
In this paper, the main work is ultrasound image segmentation, so the normal image should be removed. Due to the large gap between benign and malignant in terms of boundary and texture in appearance, combined with the previous work, this dataset was disassembled for segmentation work\cite{zhou2022laednet},\cite{zou2021robust}, this work regards benign and malignant as two fields: BUSI(Benign) and BUSI(Malignant), respectively. 
\subsubsection{Digital Database of Thyroid Ultrasound Images}
The \href{http://cimalab.intec.co/applications/thyroid/}{DDTI} dataset\cite{pedraza2015open} is an open-access resource for the scientific community.
Support for this project was provided by the National University of Colombia, CIM@LAB and IDIME (Instituto de Diagnostico Medico).
The main purpose of the dataset DDTI is to develop algorithms for applying CAD systems to the analysis of thyroid nodules, containing annotations and patient diagnostic information for manually labeled suspicious lesions using the TI-RADS classification by experienced radiologists.
The dataset consists of raw thyroid images of size 560 × 360 pixels, and the pixel labels of the tumor contour regions are given in the corresponding XML files. When performing the segmentation task, the XML tag should be converted into a mask before it can be used.
The DDTI dataset contains 390 cases with 91 normal, 52 benign, and 247 malignant categories, and some single cases have multiple view images.
It can be seen from the image examples that benign and malignant images are very close in appearance, and in the previous work \cite{ding2019automatic}, \cite{koundal2018computer}, for the segmentation task, the benign and malignant images in the dataset are mixed. Therefore, in the following text, the two sub-datasets in this dataset will be regarded as the same source domain, abbreviated as DDTI.
\subsubsection{CT2USforKidneySeg}
The \href{https://www.kaggle.com/siatsyx/ct2usforkidneyseg}{CT2USforKidneySeg} dataset is a dataset provided by the literature \cite{song2022ct2us}, which uses labeled real CT data and some ultrasound data to synthesize more ultrasound images to construct a transition dataset, and then migrate to real ultrasound images for segmentation. The dataset has a total of 4,586 samples and is simply called Kidney.
	\subsection{Data preprocessing}
\begin{figure}[!t]
	\centerline{\includegraphics[width=\columnwidth]{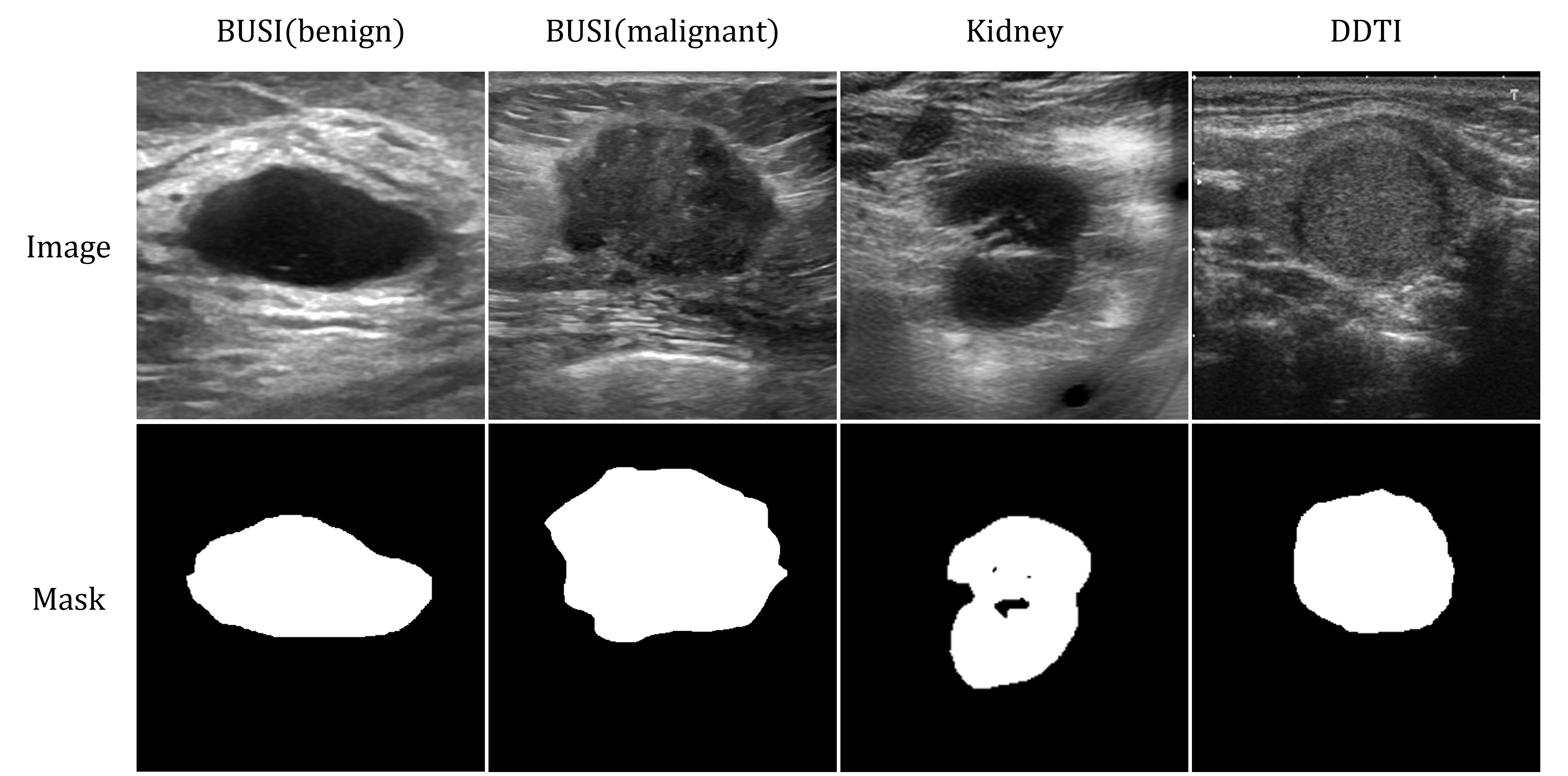}}
	\caption{An example of the processed data image of the four domains, including scaling to the same size, removing black borders, removing text, and generating a mask from an XML file.}
	\label{fig_datas}
\end{figure}

In terms of data preprocessing, due to the presence of black borders and text in some data, the performance of the algorithm will be degraded if used directly, so the black borders and text should be deleted during processing;
For the dataset, firstly, the target domain dataset is divided into training set, validation set and test set according to the ratio of $0.8:0.1:0.1$.
Since transfer learning focuses on tasks in the target domain, there is no need to divide the source domain dataset into a test set.
Due to the limitation of computing resources, the amount of data in each batch in the image segmentation task is small, so the method of group normalization \cite{wu2018group} is adopted to normalize the data to achieve a better segmentation effect.
Examples of images in the processed dataset are shown in Figure \ref{fig_datas}.
\subsection{Experimental indicators}
After each pixel passes through the network, it will output the predicted classification label, $TP$, $FP$, $TN$, $FN$ which are four kinds of results, indicate true positive, false positive,
true negative, and false negative, respectively.
In general, for the segmentation problem of each image, different test indicators can be selected according to different focuses. The most common indicator is the $IOU$ score (Intersection over Union), which can be understood as the real correct area and the prediction as correct. The $IOU$ is calculated as follows:
\begin{equation}
	IOU = \frac{{TP}}{{FP + FN}}
\end{equation}
In addition to the $IOU$ score indicator, $Precision$ and $Recall$ are two commonly used basic indicators. Among them, the $Precision$ score is the ratio of the number of correctly predicted true positive pixels to the total number of predicted positive pixels (from the perspective of the prediction result, how many predictions are accurate), focusing on finding the right ones; The $Recall$ score is the ratio of the correctly predicted true positive pixels to the total number of true positive pixels (from the perspective of true annotation, how many are recalled), focusing on finding all.
They are defined as follows:
\begin{equation}
	Precision = \frac{{TP}}{{TP + FP}}
\end{equation}
\begin{equation}
	Recall = \frac{{TP}}{{TP + FN}}
\end{equation}
However, sometimes the target task needs to comprehensively consider the $Precision$ score and the $Recall$ score, and the $F_{\beta}$ score comes into being. Its calculation method is as follows:
\begin{equation}
	\begin{aligned}
	{F_\beta } &= \left( {1 + {\beta ^2}} \right) \times \frac{{Precision \times Recall}}{{\left( {{\beta ^2} \times Precision} \right) + Recall}} \\
	&= \frac{{\left( {1 + {\beta ^2}} \right) \times TP}}{{\left( {1 + {\beta ^2}} \right) \times TP + {\beta ^2} \times FN + FP}}
	\end{aligned}
\end{equation}
Among them, $\beta$ can take different values according to the focus of the task indicator. When $\beta$ taken $1$, it is the $F_1$ score, which is the famous $Dice$ score; other values can also be taken, such as the $F_2$ score and the $F_{0.5}$ score, where $F_2$ The score indicates that recall is weighted more than precision and the $F_{0.5}$ score indicates that precision is weighted more than recall.
\par
At the pixel level of each sample, the test indicators selected in this paper are as follows, namely $IOU$, $Dice$, $F_{2}$ and $F_{0.5}$. The above test indicators are only for specific images in the data set. To reflect the situation of the entire data set, it is necessary to test all the images, and then take the mean and standard deviation as the data to prove the experiment.
\subsection{Experimental parameter settings}
By observing each data set, it can be seen that the background texture of BUSI(Benign) and BUSI(Malignant) is similar to that of the target Kidney's boundary and texture, so BUSI(Benign) can be selected as the target domain, BUSI(Malignant) as the source domain 1 and Kidney as the source domain 2;
In addition, the background textures of BUSI(Malignant) and BUSI(Benign) are similar, as well as the target demarcation lines and textures of DDTI, so BUSI(Malignant) as the target domain, BUSI(Benign) as the source domain 1 and DDTI as the source domain 2.
 In terms of experimental parameter setting, Pytorch is selected as the deep learning framework, and the random seed is set to 0 to avoid accidentality during model training, as shown in table \ref{Parameters_setting}, experiment with them separately. In order to intuitively reflect the performance change when the proportion of unlabeled data rises from $0\%$ to $90\%$ in the later experiments, we plot the trend of $IOU$, $Dice \ score$, $F_{2}$ and $F_{0.5}$ results for all experiments. 
\begin{table}[]
	\centering 
	\caption{Parameters setting}   
	\label{Parameters_setting}
\begin{tabular}{cc}
	\hline
	Items           & Parameters setting             \\ \hline
	Prediction loss & Cross-Entropy                  \\
	Total epochs    & 100                            \\
	Batch size      & 16                             \\
	Optimizer       & Root Mean Square Prop(RMSprop) \\
	Learning rate   & {[}0.001,0.0001{]}             \\ \hline
\end{tabular}
\end{table}

	\subsection{Compare experimental results and analysis}
	In terms of comparative experiments, this work selected some common multi-source transfer learning methods described above and compared them with the proposed method. At the data level, \cite{ge2014handling} is selected to assign different weights to different source domain data as $Comparison_1$. At the model level, we chose \cite{li2019multi} for multi-model domain adaptation as $Comparison_2$. In addition, we tested also assigning different weights to different source domain models according to the similarity of the dataset. Simultaneously all model outputs are tested by voting method \cite{she2022multi} as $Comparison_3$. We assign pixel-level pseudo-labels \cite{li2022dynamic} to some data, and then try to do domain adaptation as $Comparison_4$.
	\subsubsection{BUSI(Benign) Comparative Experiment Results and Analysis}
\begin{figure}[h]
		\centerline{\includegraphics[width=\columnwidth]{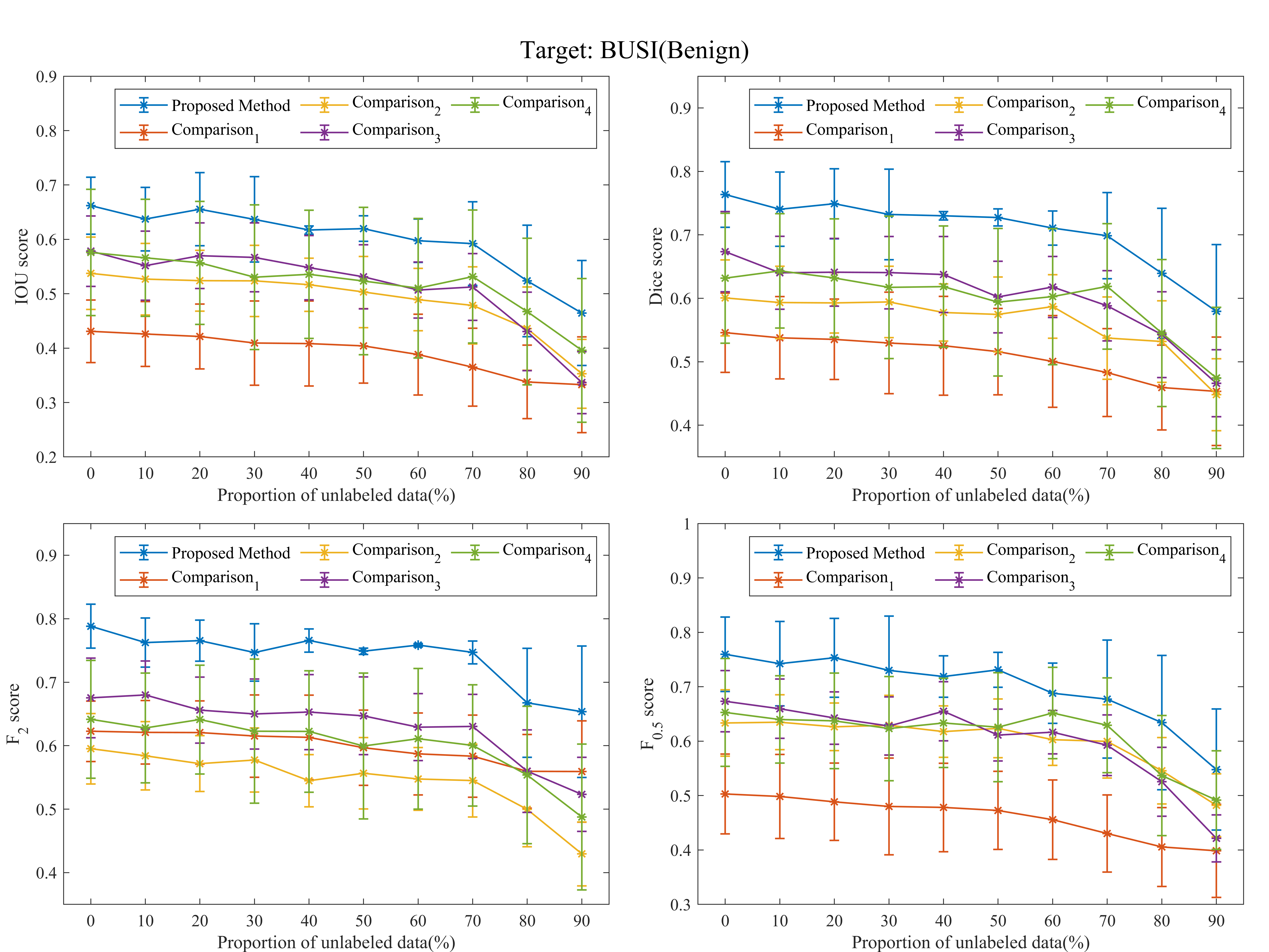}}
		\caption{The method comparison experiment using BUSI (Benign) as the target domain, and BUSI (Malignant) and CT2US as the two source domains. The curves of the four indicators of $IOU$, $Dice \ score$, $F_{2}$ and $F_{0.5}$ as a function of the proportion of unlabeled data are plotted.}
		\label{Benign_method}
	\end{figure}

	We tested the performance metrics $IOU$, $Dice \ score$, $F_{2}$ and $F_{0.5}$ on the BUSI(Benign) dataset with different proportions of unlabeled data. The results are shown in the table \ref{Comparison_Benign_score_result}. Figure \ref{Benign_method} plots the trend of all experiments in the table \ref{Comparison_Benign_score_result} as the proportion of unlabeled data increases from $0\%$ to $90\%$.
	It can be seen that among the four indicators, $Proposed \ Method$ has the best performance. The method $Comparison_1$ of assigning different weights to the data has the worst effect. We believe that this is because the overall similarity is insufficient, and simply weighting the data cannot filter out the desired features at the local feature level. The performance of $Comparison_2$ and $Comparison_3$ at the model level is close to that of the pseudo-label-based method $Comparison_4$. We believe that the multi-source transfer learning and pseudo-label methods at the model level are feasible. However, the overall similarity between domains in this work is limited, and we believe that these methods may not be well suited to handle such scenarios where there are only partial similarities between the source domain and each target domain.
\begin{table*}[thp]
	\tiny  
	\centering 
	\caption{Results of BUSI(Benign) comparative experiment}   
	\label{Comparison_Benign_score_result}
	\resizebox{\textwidth}{!}
	{
\begin{tabular}{clllllllllll}
	\hline
	\multicolumn{2}{c}{Target: BUSI(Benign)} &
	\multicolumn{10}{c}{Proportion of unlabeled data} \\
	\multicolumn{1}{c}{Score(\%)} &
	\multicolumn{1}{c}{Experiment} &
	0\% &
	10\% &
	20\% &
	30\% &
	40\% &
	50\% &
	60\% &
	70\% &
	80\% &
	90\% \\ \hline
	\multirow{5}{*}{$IoU$} &
	$Proposed \ Method$ &
	66.197±5.239 &
	63.725±5.832 &
	65.545±6.714 &
	63.68±7.845 &
	61.718±0.76 &
	61.986±2.343 &
	59.749±3.998 &
	59.216±7.682 &
	52.359±10.259 &
	46.45±9.653 \\
	&
	$Comparison_1$ &
	43.108±5.751 &
	42.594±5.963 &
	42.142±5.964 &
	40.926±7.735 &
	40.847±7.82 &
	40.407±6.848 &
	38.813±7.424 &
	36.495±7.179 &
	33.79±6.755 &
	33.274±8.798 \\
	&
	$Comparison_2$ &
	53.76±6.657 &
	52.681±6.578 &
	52.399±5.602 &
	52.357±6.544 &
	51.653±4.88 &
	50.31±6.554 &
	48.923±5.728 &
	47.848±7.089 &
	43.595±7.655 &
	35.288±6.335 \\
	&
	$Comparison_3$ &
	57.832±6.481 &
	55.157±6.365 &
	57.004±6.029 &
	56.695±6.325 &
	54.792±5.92 &
	53.124±5.889 &
	50.681±5.154 &
	51.239±6.15 &
	43.072±7.224 &
	33.659±5.692 \\
	&
	$Comparison_4$ &
	57.592±11.596 &
	56.617±10.761 &
	55.683±11.305 &
	53.04±13.31 &
	53.573±11.801 &
	52.338±13.562 &
	51.016±12.834 &
	53.166±12.22 &
	46.728±13.477 &
	39.598±13.215 \\ \hdashline[1pt/1pt]
	\multirow{5}{*}{$Dice$} &
	$Proposed \ Method$ &
	76.361±5.16 &
	74.044±5.86 &
	74.893±5.53 &
	73.212±7.127 &
	72.99±0.663 &
	72.743±1.347 &
	71.055±2.694 &
	69.868±6.797 &
	63.931±10.241 &
	57.94±10.535 \\
	&
	$Comparison_1$ &
	54.572±6.252 &
	53.769±6.487 &
	53.533±6.345 &
	52.964±7.988 &
	52.522±7.791 &
	51.586±6.817 &
	50.039±7.225 &
	48.293±6.94 &
	45.936±6.687 &
	45.332±8.546 \\
	&
	$Comparison_2$ &
	60.049±5.985 &
	59.345±5.701 &
	59.284±4.754 &
	59.429±5.627 &
	57.76±4.517 &
	57.449±5.906 &
	58.704±5.002 &
	53.722±6.496 &
	53.177±6.426 &
	44.802±5.671 \\
	&
	$Comparison_3$ &
	67.35±6.322 &
	64.013±5.745 &
	64.095±5.341 &
	64.036±5.703 &
	63.743±5.988 &
	60.194±5.632 &
	61.783±4.817 &
	58.828±5.535 &
	54.268±6.757 &
	46.602±5.274 \\
	&
	$Comparison_4$ &
	63.163±10.244 &
	64.317±9.007 &
	63.191±9.33 &
	61.722±11.222 &
	61.831±9.557 &
	59.379±11.64 &
	60.266±10.72 &
	61.873±9.888 &
	54.532±11.58 &
	47.44±11.125 \\ \hdashline[1pt/1pt]
	\multirow{5}{*}{$F_2$} &
	$Proposed \ Method$ &
	78.823±3.467 &
	76.246±3.881 &
	76.548±3.234 &
	74.679±4.515 &
	76.567±1.82 &
	74.873±0.507 &
	75.847±0.254 &
	74.691±1.802 &
	66.746±8.591 &
	65.353±10.359 \\
	&
	$Comparison_1$ &
	62.266±4.771 &
	62.113±4.99 &
	62.075±4.981 &
	61.512±6.489 &
	61.317±6.65 &
	59.674±5.919 &
	58.698±6.455 &
	58.354±6.475 &
	55.974±5.803 &
	55.927±7.985 \\
	&
	$Comparison_2$ &
	59.518±5.538 &
	58.408±5.371 &
	57.181±4.404 &
	57.742±5.04 &
	54.477±4.108 &
	55.669±5.623 &
	54.765±4.937 &
	54.513±5.743 &
	49.973±5.879 &
	42.95±5.026 \\
	&
	$Comparison_3$ &
	67.525±6.279 &
	68.001±5.358 &
	65.615±5.203 &
	64.987±5.509 &
	65.29±5.911 &
	64.706±6.129 &
	62.925±5.27 &
	63.032±5.05 &
	55.988±6.503 &
	52.341±5.831 \\
	&
	$Comparison_4$ &
	64.16±9.281 &
	62.787±8.651 &
	64.11±8.563 &
	62.289±11.36 &
	62.238±9.564 &
	59.958±11.491 &
	61.095±11.077 &
	60.038±9.562 &
	55.394±10.824 &
	48.769±11.485 \\ \hdashline[1pt/1pt]
	\multirow{5}{*}{$F_{0.5}$} &
	$Proposed \ Method$ &
	75.972±6.84 &
	74.245±7.76 &
	75.318±7.253 &
	73.013±9.98 &
	71.873±3.821 &
	73.12±3.217 &
	68.812±5.543 &
	67.723±10.848 &
	63.42±12.336 &
	54.783±11.118 \\
	&
	$Comparison_1$ &
	50.299±7.346 &
	49.832±7.726 &
	48.86±7.1 &
	48.009±8.889 &
	47.812±8.131 &
	47.273±7.187 &
	45.575±7.295 &
	43.028±7.074 &
	40.551±7.24 &
	39.842±8.557 \\
	&
	$Comparison_2$ &
	63.335±6.109 &
	63.485±5.054 &
	62.627±4.374 &
	62.921±5.498 &
	61.745±4.724 &
	62.356±5.385 &
	60.286±4.752 &
	59.952±6.739 &
	54.557±6.112 &
	48.211±5.781 \\
	&
	$Comparison_3$ &
	67.341±5.616 &
	65.968±5.453 &
	64.256±4.817 &
	62.795±5.349 &
	65.491±5.437 &
	61.127±4.751 &
	61.64±3.982 &
	59.263±5.578 &
	52.543±6.337 &
	42.132±4.334 \\
	&
	$Comparison_4$ &
	65.278±9.922 &
	63.992±8.014 &
	63.742±8.786 &
	62.306±9.572 &
	63.356±8.208 &
	62.575±10.013 &
	65.177±8.406 &
	62.919±8.697 &
	53.679±11.05 &
	49.171±9.057 \\ \hline
\end{tabular}
}
\end{table*}
	\subsubsection{BUSI(Malignant) comparative experiment results and analysis}
\begin{figure}[h]
		\centerline{\includegraphics[width=\columnwidth]{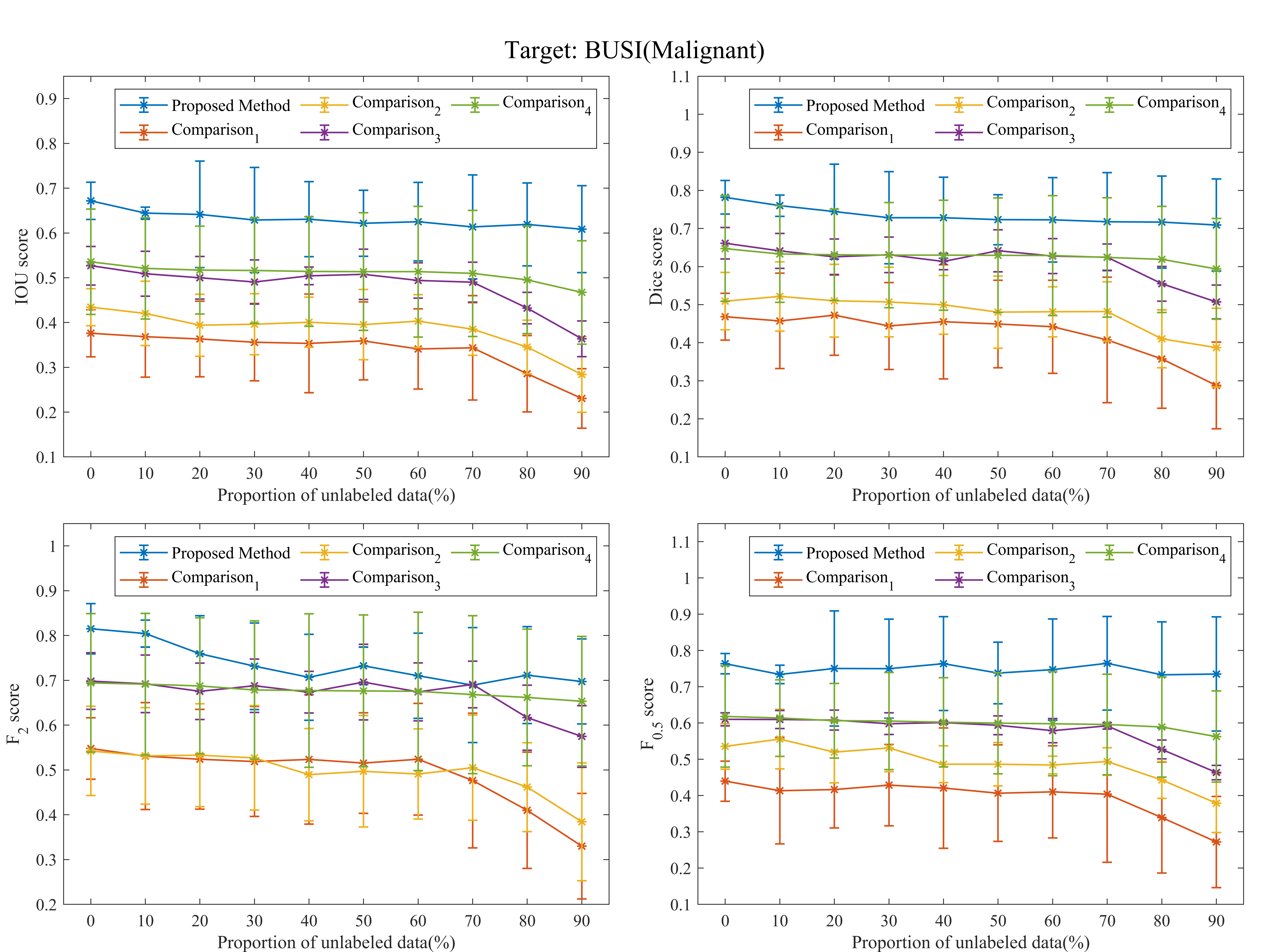}}
		\caption{The method comparison experiment with BUSI (Malignant) as the target domain, and BUSI (Benign) and DDTI as the two source domains. The curves of the four indicators of $IOU$, $Dice \ score$, $F_{2}$ and $F_{0.5}$ as a function of the proportion of unlabeled data are plotted.}
		\label{Malignant_method}
	\end{figure}

	Similar to the comparison experiment of BUSI (Benign), the figure \ref{Malignant_method} draws the trend of the table \ref{Malignant_method}. Also similar to the comparison experiment of BUSI (Benign), $Proposed \ Method$ has the best performance. The method $Comparison_1$ that assigns different weights to the data is the worst. It is worth noting that for this data set, $Comparison_1$, $Comparison_2$ and $Comparison_3$ have a significant performance drop when the unlabeled data reaches about $70\%$. And the performance drop of $Comparison_4$ and $Proposed \ Method$ has little change. We believe this is because these methods are better able to extract the key features needed to cover the target domain from the selected source domain than the experiments on the target domain BUSI (Malignant) compared to the experiments on the target domain BUSI (Benign).
	\begin{table}[thp]
	\centering 
	\caption{Results of BUSI(Malignant) comparative experiment}   
	\label{Comparison_Malignant_score}
	\resizebox{\textwidth}{!}
	{
		\begin{tabular}{llllllllllll}
			\hline
			\multicolumn{2}{c}{Target: BUSI(Malignant)} &
			\multicolumn{10}{c}{Proportion of   unlabeled data} \\
			\multicolumn{1}{c}{Score(\%)} &
			\multicolumn{1}{c}{Experiment} &
			0\% &
			10\% &
			20\% &
			30\% &
			40\% &
			50\% &
			60\% &
			70\% &
			80\% &
			90\% \\ \hline
			\multirow{5}{*}{$IoU$} &
			$Proposed \ Method$ &
			67.177±4.149 &
			64.422±1.358 &
			64.154±11.891 &
			62.906±11.746 &
			63.08±8.387 &
			62.168±7.378 &
			62.521±8.765 &
			61.352±11.598 &
			61.907±9.248 &
			60.841±9.704 \\
			&
			$Comparison_1$ &
			37.593±5.256 &
			36.842±9.021 &
			36.31±8.438 &
			35.619±8.623 &
			35.351±11.015 &
			35.907±8.704 &
			34.102±8.939 &
			34.353±11.639 &
			28.559±8.519 &
			23.062±6.642 \\
			&
			$Comparison_2$ &
			43.432±4.141 &
			42.044±7.176 &
			39.392±6.903 &
			39.625±6.796 &
			40.075±5.594 &
			39.537±7.833 &
			40.326±5.92 &
			38.531±5.861 &
			34.549±5.982 &
			28.392±8.439 \\
			&
			$Comparison_3$ &
			52.676±4.324 &
			50.895±4.993 &
			49.981±4.757 &
			49.06±4.922 &
			50.421±1.975 &
			50.762±5.608 &
			49.395±3.955 &
			48.999±4.452 &
			43.24±3.516 &
			36.366±3.973 \\
			&
			$Comparison_4$ &
			53.573±11.761 &
			52.092±11.317 &
			51.696±9.811 &
			51.625±11.798 &
			51.391±12.25 &
			51.383±13.144 &
			51.352±14.601 &
			50.975±14.081 &
			49.537±12.015 &
			46.731±11.539 \\  \hdashline[1pt/1pt]
			\multirow{5}{*}{$Dice$} &
			$Proposed \ Method$ &
			78.194±4.405 &
			75.988±2.793 &
			74.431±12.49 &
			72.827±12.112 &
			72.83±10.631 &
			72.317±6.573 &
			72.262±11.084 &
			71.795±12.875 &
			71.676±12.112 &
			70.921±12.085 \\
			&
			$Comparison_1$ &
			46.818±6.131 &
			45.726±12.52 &
			47.218±10.543 &
			44.39±11.417 &
			45.505±14.991 &
			44.893±11.496 &
			44.177±12.207 &
			40.72±16.461 &
			35.712±12.937 &
			28.759±11.397 \\
			&
			$Comparison_2$ &
			50.915±7.537 &
			52.148±9.1 &
			51.018±9.564 &
			50.679±9.137 &
			49.93±7.704 &
			48.024±9.492 &
			48.131±6.579 &
			48.202±7.78 &
			41.009±7.594 &
			38.709±10.348 \\
			&
			$Comparison_3$ &
			66.127±4.124 &
			64.121±4.575 &
			62.568±4.679 &
			63.095±4.666 &
			61.318±2.136 &
			64.137±5.519 &
			62.755±4.574 &
			62.458±3.45 &
			55.474±4.553 &
			50.701±4.439 \\
			&
			$Comparison_4$ &
			64.733±13.901 &
			63.324±12.703 &
			63.124±11.993 &
			62.995±13.817 &
			62.99±14.435 &
			62.974±15.059 &
			62.884±15.746 &
			62.402±15.665 &
			61.875±13.923 &
			59.391±13.244 \\  \hdashline[1pt/1pt]
			\multirow{5}{*}{$F_2$} &
			$Proposed \ Method$ &
			81.505±5.605 &
			80.462±2.995 &
			75.952±8.46 &
			73.144±9.648 &
			70.675±9.604 &
			73.262±4.19 &
			71.033±9.51 &
			68.938±12.816 &
			71.17±10.811 &
			69.753±9.483 \\
			&
			$Comparison_1$ &
			54.79±6.844 &
			53.094±11.933 &
			52.401±11.134 &
			51.895±12.274 &
			52.349±14.449 &
			51.544±11.228 &
			52.401±12.45 &
			47.635±15.017 &
			40.999±12.988 &
			32.999±11.777 \\
			&
			$Comparison_2$ &
			54.246±9.955 &
			53.143±10.782 &
			53.282±11.48 &
			52.714±11.683 &
			48.947±10.289 &
			49.714±12.451 &
			49.091±10.066 &
			50.503±11.732 &
			46.165±9.914 &
			38.435±13.155 \\
			&
			$Comparison_3$ &
			69.839±6.309 &
			69.238±6.422 &
			67.558±6.288 &
			68.798±5.94 &
			67.355±4.66 &
			69.608±8.443 &
			67.429±6.47 &
			69.077±5.219 &
			61.681±7.258 &
			57.478±6.871 \\
			&
			$Comparison_4$ &
			69.434±15.443 &
			69.14±15.786 &
			68.76±15.184 &
			67.843±15.447 &
			67.721±17.138 &
			67.645±16.934 &
			67.534±17.682 &
			66.804±17.62 &
			66.186±15.261 &
			65.334±14.453 \\  \hdashline[1pt/1pt]
			\multirow{5}{*}{$F_{0.5}$} &
			$Proposed \ Method$ &
			76.354±2.787 &
			73.403±2.529 &
			75.043±15.874 &
			74.982±13.667 &
			76.376±12.921 &
			73.777±8.488 &
			74.74±13.963 &
			76.469±12.901 &
			73.277±14.622 &
			73.53±15.722 \\
			&
			$Comparison_1$ &
			43.969±5.516 &
			41.343±14.682 &
			41.65±10.557 &
			42.879±11.205 &
			42.062±16.596 &
			40.657±13.302 &
			40.997±12.696 &
			40.39±18.799 &
			33.929±15.28 &
			27.198±12.571 \\
			&
			$Comparison_2$ &
			53.533±6.29 &
			55.554±8.192 &
			51.971±8.45 &
			53.138±6.551 &
			48.662±5.082 &
			48.647±5.993 &
			48.431±2.454 &
			49.376±3.808 &
			44.301±5.102 &
			37.92±8.133 \\
			&
			$Comparison_3$ &
			61.035±1.801 &
			60.984±2.486 &
			60.823±2.753 &
			59.826±3 &
			60.125±0.427 &
			59.379±2.581 &
			57.891±3.352 &
			59.203±0.892 &
			52.71±2.576 &
			46.343±1.995 \\
			&
			$Comparison_4$ &
			61.794±13.973 &
			61.373±10.565 &
			60.634±10.303 &
			60.54±13.397 &
			60.207±12.311 &
			59.938±13.94 &
			59.804±14.32 &
			59.575±13.86 &
			58.907±13.787 &
			56.272±12.56 \\ \hline
		\end{tabular}
	}
\end{table}
\subsection{Ablation experiment Results and Analysis}
Since this paper uses adversarial transfer learning to solve the ultrasound image segmentation problem in the absence of labeled data, the core idea is to use adversarial transfer learning to extract common features between the source and target domains;
However, considering that the common features provided by a single source domain are always limited, multi-source transfer learning is adopted to supplement the available common features.
Therefore, this paper needs to verify the effectiveness of adversarial transfer learning and multi-source learning respectively, as well as multi-source adversarial transfer learning, so the selected experiment types are shown in table \ref{experiments_setting}.\begin{table}[]
	\caption{Experiments settings}   
	\label{experiments_setting}
	\begin{tabular}{cccccc}
		\hline
		\multicolumn{2}{c}{Name}                  & Target & Source 1 & Source 2 & Method                        \\ \hline
		\multirow{3}{*}{Adversarial:} & 2 Sources & \checkmark      & \checkmark        & \checkmark        & ATL$^a$ \\
		& Source 1  & \checkmark      & \checkmark        &          & ATL \\
		& Source 2  & \checkmark      &          & \checkmark        & ATL \\ \hdashline[1pt/1pt]
		\multirow{3}{*}{Ablation:}    & 2 Sources & \checkmark      & \checkmark        & \checkmark        & MTL$^b$           \\
		& Source 1  & \checkmark      & \checkmark        &          & MTL           \\
		& Source 2  & \checkmark      &          & \checkmark        & MTL           \\ \hdashline[1pt/1pt]
		\multicolumn{2}{c}{Only Target}           & \checkmark      &          &          & DL$^c$                 \\ \hline
	\end{tabular}\\
\footnotesize{$^a$ ATL $=$ Adversarial Transfer Learning}\\
\footnotesize{$^b$ MTL $=$ Multi-task Learning}\\
\footnotesize{$^c$ DL $=$ Deep Learning}
\end{table} 
\subsubsection{BUSI(Benign) Ablation experiment Results and Analysis}
\begin{figure}[!t]
	\centerline{\includegraphics[width=\columnwidth]{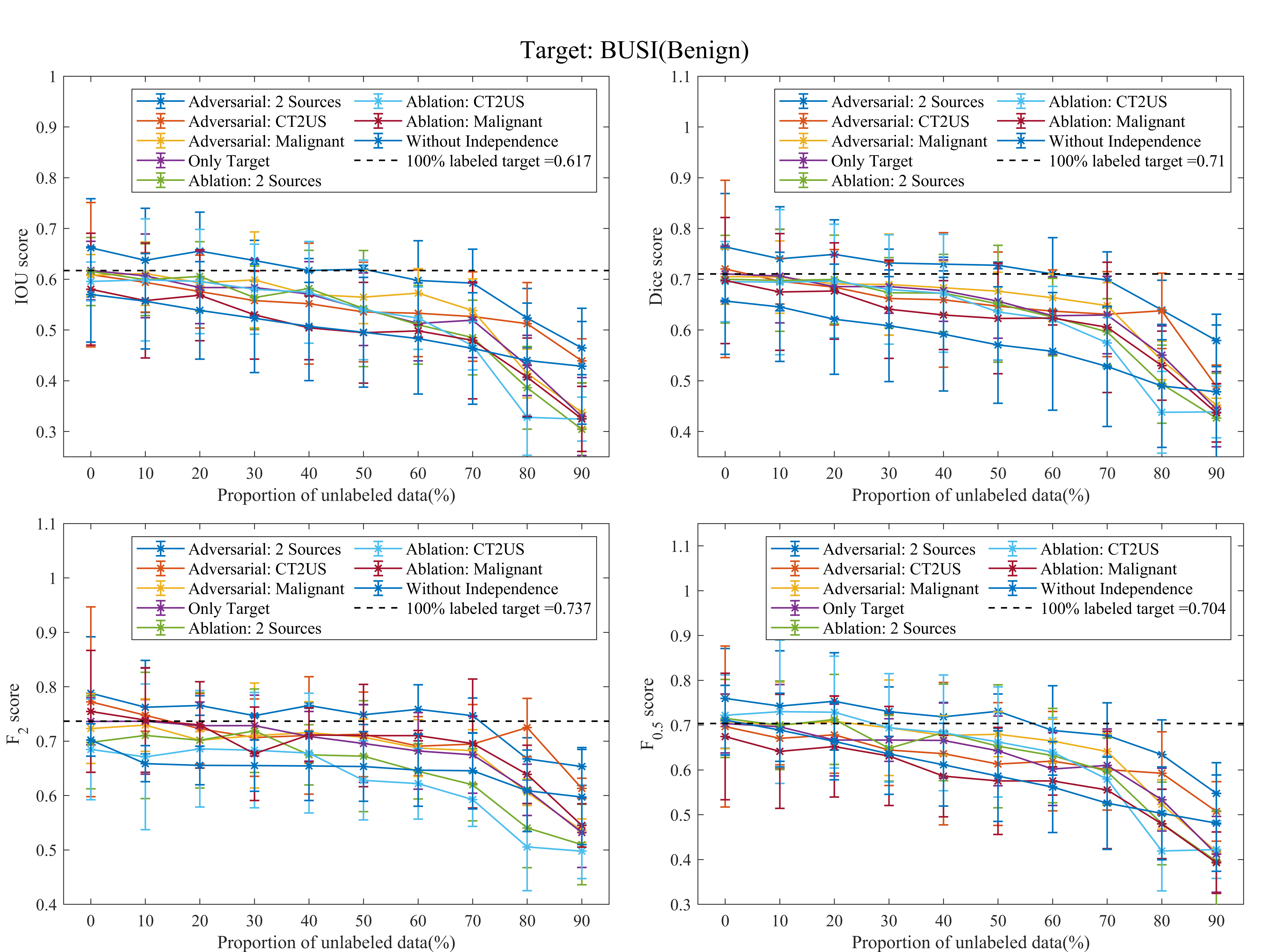}}
	\caption{Ablation experiments with BUSI (Benign) as the target domain and BUSI (Malignant) and CT2US as the two source domains. The curves of the four indicators of $IOU$, $Dice \ score$, $F_{2}$ and $F_{0.5}$ as a function of the proportion of unlabeled data are plotted.}
	\label{Benign}
\end{figure}
\begin{table*}[thp]
	\centering 
	\caption{BUSI(Benign) ablation experiment results}   
	\label{Benign_score_result}
	\resizebox{\textwidth}{!}
	{
	\begin{tabular}{cclllllllllll}
		\hline
		\multicolumn{3}{c}{Target: BUSI(Benign)} &
		\multicolumn{10}{c}{The proportion of unlabeled data} \\
		\multicolumn{1}{c}{Score(\%)} &
		\multicolumn{2}{c}{Experiment} &
		0\% &
		10\% &
		20\% &
		30\% &
		40\% &
		50\% &
		60\% &
		70\% &
		80\% &
		90\% \\ \hline
		\multirow{8}{*}{$IoU$} &
		\multirow{3}{*}{Adversarial:} &
		2 Sources &
		66.197±5.239 &
		63.725±5.832 &
		65.545±6.714 &
		63.68±7.845 &
		61.718±0.76 &
		61.986±2.343 &
		59.749±3.998 &
		59.216±7.682 &
		52.359±10.259 &
		46.45±9.653 \\
		&
		&
		CT2US &
		60.9±4.367 &
		59.386±8.085 &
		57.574±8.791 &
		55.808±8.522 &
		55.198±9.812 &
		53.548±11.906 &
		53.289±5.716 &
		52.654±7.172 &
		51.252±5.856 &
		43.914±14.226 \\
		&
		&
		Malignant &
		60.766±3.009 &
		61.109±4.809 &
		59.299±6.232 &
		59.884±4.831 &
		57.004±5.233 &
		56.484±6.479 &
		57.281±9.46 &
		53.816±0.656 &
		41.43±6.25 &
		33.782±4.118 \\
		&
		\multirow{3}{*}{Ablation:} &
		2 Sources &
		61.713±7.668 &
		60.669±5.94 &
		58.368±7.383 &
		58.368±7.383 &
		57.066±7.141 &
		54.087±6.399 &
		51.297±5.606 &
		51.94±7.11 &
		43.023±8.258 &
		32.946±5.767 \\
		&
		&
		CT2US &
		61.524±9.213 &
		59.918±8.133 &
		60.541±7.359 &
		56.406±7.748 &
		58.179±11.427 &
		54.224±7.5 &
		51.037±6.199 &
		48.505±6.852 &
		38.609±7.062 &
		30.407±6.697 \\
		&
		&
		Malignant &
		59.571±4.337 &
		59.872±7.471 &
		59.555±4.771 &
		58.028±6.15 &
		57.43±9.812 &
		53.95±10.013 &
		52.334±8.863 &
		46.902±10.273 &
		32.808±12.033 &
		32.444±3.832 \\
		&
		\multicolumn{2}{c}{Only Target} &
		58.04±6.394 &
		55.809±7.7 &
		56.863±11.512 &
		52.98±1.232 &
		50.426±9.927 &
		49.471±6.288 &
		49.81±8.704 &
		47.962±8.954 &
		40.707±11.326 &
		32.487±11.031 \\
		&
		\multicolumn{2}{c}{Without Independence} &
		56.998±11.417 &
		55.67±11.294  &
		53.867±10.983 &
		52.295±10.933 &
		50.717±10.808 &
		49.551±10.704 &
		48.319±10.69 &
		46.373±9.589 &
		44.015±9.388 &
		42.873±9.373 \\ \hdashline[1pt/1pt]
		\multirow{8}{*}{$Dice$} &
		\multirow{3}{*}{Adversarial:} &
		2 Sources &
		76.361±5.16 &
		74.044±5.86 &
		74.893±5.53 &
		73.212±7.127 &
		72.99±0.663 &
		72.743±1.347 &
		71.055±2.694 &
		69.868±6.797 &
		63.931±10.241 &
		57.94±10.535 \\
		&
		&
		CT2US &
		72.029±4.178 &
		69.617±7.424 &
		68.426±8.388 &
		66.193±8.141 &
		65.927±10.667 &
		64.686±13.259 &
		63.71±7.218 &
		63.125±7.438 &
		63.783±4.518 &
		48.892±17.473 \\
		&
		&
		Malignant &
		70.512±1.358 &
		70.431±3.775 &
		69.07±4.516 &
		68.951±4.697 &
		68.318±3.734 &
		67.659±5.143 &
		66.345±9.944 &
		64.822±0.878 &
		54.006±7.13 &
		45.172±5.432 \\
		&
		\multirow{3}{*}{Ablation:} &
		2 Sources &
		71.045±7.403 &
		70.628±5.748 &
		68.554±7.636 &
		68.554±7.636 &
		67.754±7.277 &
		65.671±6.056 &
		62.817±5.248 &
		62.944±6.246 &
		55.066±9.22 &
		44.392±5.493 \\
		&
		&
		CT2US &
		70.004±8.821 &
		69.779±7.679 &
		69.946±6.548 &
		67.325±7.629 &
		67.262±11.517 &
		65.166±7.645 &
		62.534±6.945 &
		59.625±8.743 &
		49.328±10.039 &
		42.636±8.623 \\
		&
		&
		Malignant &
		69.539±5.111 &
		69.421±8.053 &
		69.644±4.991 &
		67.98±6.493 &
		67.23±9.454 &
		63.579±11.596 &
		62.1±10.779 &
		57.49±11.184 &
		43.797±14.284 &
		43.878±7.893 \\
		&
		\multicolumn{2}{c}{Only Target} &
		69.75±5.746 &
		67.497±6.813 &
		67.666±12.842 &
		64.074±1.356 &
		62.946±10.883 &
		62.28±6.79 &
		62.358±9.671 &
		60.523±9.489 &
		52.978±11.481 &
		43.658±12.421 \\
		&
		\multicolumn{2}{c}{Without Independence} &
		65.721±13.133 &
		64.551±12.131 &
		62.124±11.833 &
		60.851±11.592 &
		59.178±11.481 &
		57.049±11.201 &
		55.804±11.003 &
		52.817±10.84  &
		48.984±10.754 &
		47.851±10.514 \\ \hdashline[1pt/1pt]
		\multirow{8}{*}{$F_2$} &
		\multirow{3}{*}{Adversarial:} &
		2 Sources &
		78.823±3.467 &
		76.246±3.881 &
		76.548±3.234 &
		74.679±4.515 &
		76.567±1.82 &
		74.873±0.507 &
		75.847±0.254 &
		74.691±1.802 &
		66.746±8.591 &
		65.353±10.359 \\
		&
		&
		CT2US &
		77.229±1.863 &
		74.738±5.363 &
		72.279±7.284 &
		70.652±5.419 &
		71.048±7.772 &
		71.249±10.777 &
		69.071±7.102 &
		69.448±6.463 &
		72.49±2.906 &
		61.324±17.439 \\
		&
		&
		Malignant &
		72.304±2.178 &
		72.944±2.452 &
		70.176±3.826 &
		71.024±5.218 &
		71.569±3.323 &
		70.725±5.685 &
		68.714±9.651 &
		68.274±1.604 &
		60.626±4.828 &
		53.543±6.386 \\
		&
		\multirow{3}{*}{Ablation:} &
		2 Sources &
		73.627±6.436 &
		73.672±4.704 &
		72.854±7.039 &
		72.854±7.039 &
		70.891±7.153 &
		69.567±4.571 &
		68.204±5.587 &
		67.489±4.343 &
		61.042±9.755 &
		53.22±4.318 \\
		&
		&
		CT2US &
		69.781±7.391 &
		71.053±7.329 &
		70.153±6.624 &
		71.903±5.074 &
		67.479±10.195 &
		67.256±5.543 &
		64.443±7.666 &
		61.986±8.773 &
		54.064±11.59 &
		50.987±8.537 \\
		&
		&
		Malignant &
		68.489±5.029 &
		67.116±8.016 &
		68.585±4.962 &
		68.346±6.512 &
		67.819±7.273 &
		62.812±11.005 &
		62.184±10.6 &
		59.285±10.703 &
		50.548±13.376 &
		49.784±9.271 \\
		&
		\multicolumn{2}{c}{Only Target} &
		75.475±3.98 &
		73.886±5.365 &
		72.992±11.849 &
		67.685±0.94 &
		71.158±9.387 &
		71.029±4.956 &
		71.013±8.585 &
		69.582±7.923 &
		63.894±9.601 &
		54.507±11.205 \\ 
		&
		\multicolumn{2}{c}{Without Independence} &
		70.217±8.743 &
		65.878±7.486 &
		65.535±7.002 &
		65.507±6.598 &
		65.449±6.404 &
		65.348±6.372 &
		64.65±4.727 &
		64.556±3.547 &
		60.881±3.313 &
		59.739±2.993 \\ \hdashline[1pt/1pt]
		\multirow{8}{*}{$F_{0.5}$} &
		\multirow{3}{*}{Adversarial:} &
		2 Sources &
		75.972±6.84 &
		74.245±7.76 &
		75.318±7.253 &
		73.013±9.98 &
		71.873±3.821 &
		73.12±3.217 &
		68.812±5.543 &
		67.723±10.848 &
		63.42±12.336 &
		54.783±11.118 \\
		&
		&
		CT2US &
		69.693±6.64 &
		67.081±9.276 &
		67.841±9.029 &
		64.434±11.084 &
		63.633±13.723 &
		61.322±15.885 &
		61.959±7.908 &
		60.09±8.591 &
		59.251±5.939 &
		50.749±17.953 \\
		&
		&
		Malignant &
		70.807±0.332 &
		70.052±5.501 &
		70.839±4.008 &
		69.427±4.846 &
		67.699±5.03 &
		67.962±4.413 &
		66.462±10.663 &
		64.125±0.001 &
		52.315±9.516 &
		41.767±6.04 \\
		&
		\multirow{3}{*}{Ablation:} &
		2 Sources &
		70.369±8.475 &
		69.636±7.013 &
		66.675±8.133 &
		66.675±8.133 &
		66.587±7.776 &
		64.265±8.373 &
		60.2±4.746 &
		61.036±8.014 &
		53.37±9.423 &
		41.174±6.578 \\
		&
		&
		CT2US &
		71.507±11.254 &
		69.974±9.238 &
		71.272±7.49 &
		64.835±10.527 &
		68.36±13.732 &
		65.285±10.771 &
		63.195±7.303 &
		59.653±10.046 &
		48.061±9.884 &
		39.637±8.723 \\
		&
		&
		Malignant &
		72.219±6.427 &
		73.007±8.902 &
		72.899±5.157 &
		69.409±7.715 &
		68.297±12.298 &
		66.239±12.913 &
		63.986±12.091 &
		57.957±12.48 &
		41.928±15.993 &
		42.205±8.987 \\
		&
		\multicolumn{2}{c}{Only Target} &
		67.453±6.851 &
		64.152±7.759 &
		65.232±13.101 &
		63.128±3.133 &
		58.635±11.935 &
		57.549±9.102 &
		57.538±11.054 &
		55.545±11.279 &
		47.949±12.71 &
		39.303±14.107 \\
		&
		\multicolumn{2}{c}{Without Independence} &
		71.128±10.752 &
		68.903±10.367 &
		66.358±10.296 &
		63.6±10.134 &
		61.167±10.107 &
		58.606±9.233 &
		56.172±9.061 &
		52.572±8.567 &
		50.315±8.241 &
		48.142±7.748 \\ \hline
	\end{tabular}
}
\end{table*}
Figure\ref{Benign} plots the trend of the test results in the table \ref{Benign_score_result} in the proportion of unlabeled data from $0\%$ to $90\%$. 
The multi-source adversarial transfer learning achieved the best results. Except for the adversarial transfer learning of BUSI Malignant as the source domain, the other experimental results are relatively close and not ideal.
In addition, when using the multi-source adversarial transfer learning method, only $40\%$ of all the data can be used to achieve the target domain effect of using the $100\%$ label under the same conditions.
The effect of multi-source adversarial transfer learning is also better if the target domain has a larger proportion of labeled data.
It can be seen that for deep learning that only uses the target domain, when the proportion of labeled data is insufficient, the performance is seriously degraded, overfitting occurs, and the performance on the test set is not good enough.
The effect of adversarial transfer learning using two source domains is significantly better than using only the target domain and multi-task learning ablation experiments, indicating that the common features that can be extracted through the adversarial structure are more conducive to the improvement of the target domain task.
The comparative experiments of adversarial transfer learning with two separate source domains are weaker than multi-source adversarial transfer learning at all times, and it can be considered that multiple source domains provide more information than a single source domain.
After deleting the multi-source domain independent policy, you can see a significant drop in performance. We think this is because after removing this strategy, there is no way to guarantee that the data from the target domain in each sub-batch is completely consistent, resulting in poor fusion effect. 
\subsubsection{BUSI(Malignant) Ablation experiment Results and Analysis}
\begin{figure}[!t]
	\centerline{\includegraphics[width=\columnwidth]{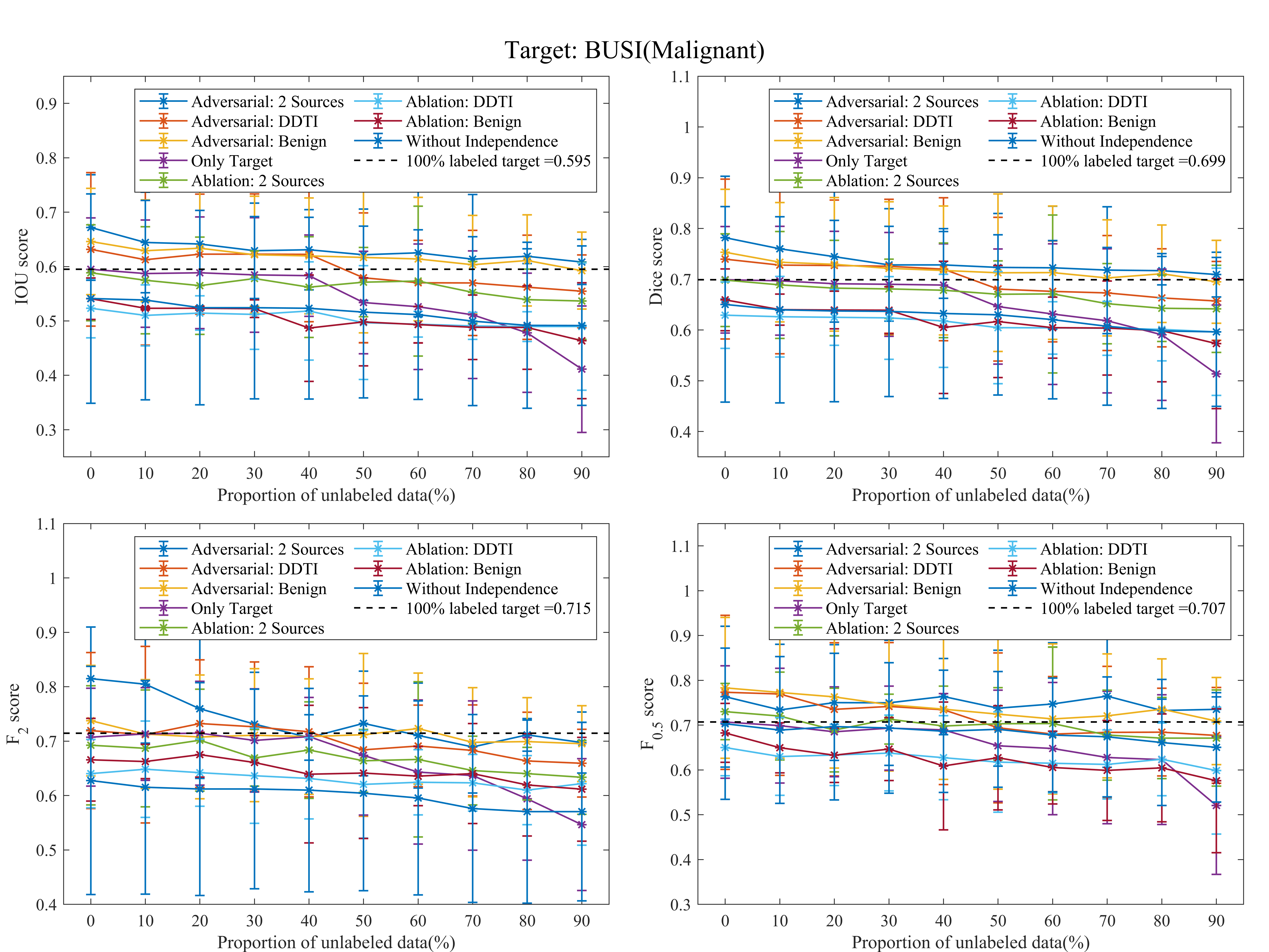}}
	\caption{Ablation experiments with BUSI (Malignant) as the target domain and BUSI (Benign) and DDTI as the two source domains. The curves of the four indicators of $IOU$, $Dice \ score$, $F_{2}$ and $F_{0.5}$ as a function of the proportion of unlabeled data are plotted.}
	\label{malignant}
\end{figure}
BUSI(Malignant)'s specific experimental results are shown in Table \ref{Malignant_score_result}. Figure \ref{malignant} plots the trend of the test results of $IOU$, $Dice$, $F_{2}$ and $F_{0.5}$ for all experiments in table \ref{experiments_setting} as the proportion of unlabeled data increases.
\begin{table*}[thp]
	\tiny  
	\centering 
	\caption{BUSI(Malignant) ablation experiment results}   
	\label{Malignant_score_result}
	\resizebox{\textwidth}{!}
	{
	\begin{tabular}{lllllllllllll}
		\hline
		\multicolumn{3}{c}{Target: BUSI(Malignant)} &
		\multicolumn{10}{c}{Proportion of unlabeled data} \\
		\multicolumn{1}{c}{Score(\%)} &
		\multicolumn{2}{c}{Experiment} &
		0\% &
		10\% &
		20\% &
		30\% &
		40\% &
		50\% &
		60\% &
		70\% &
		80\% &
		90\% \\ \hline
		\multirow{8}{*}{$IoU$} &
		\multirow{3}{*}{Adversarial:} &
		2 Sources &
		67.177±4.149 &
		64.422±1.358 &
		64.154±11.891 &
		62.906±11.746 &
		63.08±8.387 &
		62.168±7.378 &
		62.521±8.765 &
		61.352±11.598 &
		61.907±9.248 &
		60.841±9.704 \\
		&
		&
		DDTI &
		63.165±6.695 &
		61.251±9.591 &
		62.271±9.665 &
		62.262±7.773 &
		62.236±11.934 &
		57.941±12.381 &
		57.038±11.113 &
		56.991±11.051 &
		56.193±15.708 &
		55.457±14.107 \\
		&
		&
		Benign &
		64.598±7.075 &
		62.913±8.386 &
		63.384±9.065 &
		62.18±11.349 &
		61.96±13.866 &
		61.675±10.65 &
		61.39±10.747 &
		60.34±10.632 &
		61.123±9.346 &
		59.254±9.781 \\
		&
		\multirow{3}{*}{Ablation:} &
		2 Sources &
		59.506±11.618 &
		58.708±10.952 &
		58.857±11.732 &
		58.464±11.545 &
		58.321±9.413 &
		53.39±7.493 &
		52.599±10.53 &
		51.145±10.278 &
		47.825±9.861 &
		41.127±9.451 \\
		&
		&
		DDTI &
		58.856±7.066 &
		57.481±5.611 &
		56.481±5.674 &
		57.79±13.778 &
		56.222±6.382 &
		57.142±9.27 &
		57.331±5.102 &
		55.249±8.938 &
		53.908±9.847 &
		53.661±8.835 \\
		&
		&
		Benign &
		52.313±11.688 &
		51.021±2.728 &
		51.455±2.487 &
		51.203±2.351 &
		51.835±10.453 &
		49.698±9.041 &
		49.402±6.443 &
		49.094±3.154 &
		48.962±5.615 &
		48.962±5.42 \\
		&
		\multicolumn{2}{c}{Only Target} &
		54.097±10.653 &
		52.338±7.672 &
		52.343±5.942 &
		52.267±3.344 &
		48.682±8.079 &
		49.807±9.781 &
		49.328±1.63 &
		48.833±0.349 &
		48.787±0.282 &
		46.364±3.822 \\
		&
		\multicolumn{2}{c}{Without Independence} &
		54.112±14.193 &
		53.83±14.282 &
		52.45±14.482 &
		52.443±14.836 &
		52.344±15.467 &
		51.632±15.842 &
		51.156±16.09 &
		49.974±16.464 &
		49.196±17.36 &
		49.136±17.906 \\ \hdashline[1pt/1pt]
		\multirow{8}{*}{$Dice$} &
		\multirow{3}{*}{Adversarial:} &
		2 Sources &
		78.194±4.405 &
		75.988±2.793 &
		74.431±12.49 &
		72.827±12.112 &
		72.83±10.631 &
		72.317±6.573 &
		72.262±11.084 &
		71.795±12.875 &
		71.676±12.112 &
		70.921±12.085 \\
		&
		&
		DDTI &
		73.963±7.768 &
		72.779±9.66 &
		72.698±11.326 &
		72.552±9.944 &
		71.989±14.152 &
		68.045±14.082 &
		67.622±13.183 &
		67.286±12.915 &
		66.333±17.451 &
		65.733±15.75 \\
		&
		&
		Benign &
		75.304±8.156 &
		73.343±9.607 &
		72.936±11.44 &
		72.132±13.136 &
		71.73±15.526 &
		71.285±12.687 &
		71.304±13.146 &
		70.271±13.168 &
		71.092±11.758 &
		69.502±12.404 \\
		&
		\multirow{3}{*}{Ablation:} &
		2 Sources &
		69.903±13.569 &
		69.702±12.931 &
		69.134±14.195 &
		68.993±13.866 &
		68.847±11.36 &
		64.62±8.125 &
		63.132±10.22 &
		61.819±10.293 &
		59.057±10.702 &
		51.343±10.5 \\
		&
		&
		DDTI &
		69.833±8.58 &
		68.875±6.541 &
		68.257±7.901 &
		68.108±15.562 &
		67.825±6.59 &
		67.03±9.312 &
		67.098±5.851 &
		65.201±9.399 &
		64.287±10.539 &
		64.181±9.134 \\
		&
		&
		Benign &
		62.917±12.537 &
		62.617±6.179 &
		62.485±5.311 &
		62.376±5.162 &
		61.8±11.04 &
		60.477±9.169 &
		60.414±8.157 &
		60.34±5.499 &
		60.107±7.955 &
		59.643±6.544 \\
		&
		\multicolumn{2}{c}{Only Target} &
		65.957±12.825 &
		64.012±10.053 &
		63.93±9.262 &
		63.891±6.013 &
		60.512±11.051 &
		61.668±13.015 &
		60.463±4.622 &
		60.371±3.714 &
		59.854±3.065 &
		57.33±6.084 \\
		&
		\multicolumn{2}{c}{Without Independence} &
		59.629±14.656 &
		59.803±15.257 &
		60.72±15.526 &
		62.041±15.591 &
		62.987±15.789 &
		63.247±16.719 &
		63.667±16.768 &
		63.736±17.875 &
		63.98±18.335 &
		65.039±19.259 \\ \hdashline[1pt/1pt]
		\multirow{8}{*}{$F_2$} &
		\multirow{3}{*}{Adversarial:} &
		2 Sources &
		81.505±5.605 &
		80.462±2.995 &
		75.952±8.46 &
		73.144±9.648 &
		70.675±9.604 &
		73.262±4.19 &
		71.033±9.51 &
		68.938±12.816 &
		71.17±10.811 &
		69.753±9.483 \\
		&
		&
		DDTI &
		71.949±6.222 &
		71.189±8.935 &
		73.234±8.371 &
		72.646±7.483 &
		71.723±12.234 &
		68.421±11.957 &
		69.125±11.9 &
		68.29±11.727 &
		66.376±16.237 &
		65.945±14.321 \\
		&
		&
		Benign &
		73.764±6.965 &
		71.308±8.051 &
		70.78±10.024 &
		71.09±10.217 &
		70.833±14.926 &
		71.185±10.608 &
		72.305±12.221 &
		69.793±11.379 &
		69.939±9.999 &
		69.538±10.152 \\
		&
		\multirow{3}{*}{Ablation:} &
		2 Sources &
		70.719±12.12 &
		71.329±11.28 &
		71.459±13.714 &
		70.131±13.234 &
		70.84±10.949 &
		67.377±7.209 &
		64.324±9.486 &
		63.668±9.549 &
		59.399±8.52 &
		54.66±8.996 \\
		&
		&
		DDTI &
		69.235±6.781 &
		68.682±4.824 &
		70.184±6.314 &
		66.932±14.267 &
		68.366±5.769 &
		66.388±8.867 &
		66.647±5.196 &
		64.607±9.356 &
		64.028±10.739 &
		63.352±10.957 \\
		&
		&
		Benign &
		64.036±11.313 &
		64.855±6.353 &
		64.219±5.206 &
		63.637±6.001 &
		63.162±9.89 &
		62.07±7.453 &
		62.453±8.748 &
		62.353±6.171 &
		61.01±8.854 &
		62.186±6.28 \\
		&
		\multicolumn{2}{c}{Only Target} &
		66.578±9.54 &
		66.268±9.403 &
		67.529±9.179 &
		66.086±5.426 &
		63.938±12.015 &
		64.142±12.636 &
		63.58±4.94 &
		64.044±4.116 &
		61.968±3.157 &
		61.171±7.598 \\
		&
		\multicolumn{2}{c}{Without Independence} &
		57.057±16.401 &
		57.036±16.82 &
		57.621±17.266 &
		59.569±17.833 &
		60.425±17.899 &
		60.998±18.684 &
		61.224±18.336 &
		61.206±19.556 &
		61.512±19.635 &
		62.765±20.967 \\ \hdashline[1pt/1pt]
		\multirow{8}{*}{$F_{0.5}$} &
		\multirow{3}{*}{Adversarial:} &
		2 Sources &
		76.354±2.787 &
		73.403±2.529 &
		75.043±15.874 &
		74.982±13.667 &
		76.376±12.921 &
		73.777±8.488 &
		74.74±13.963 &
		76.469±12.901 &
		73.277±14.622 &
		73.53±15.722 \\
		&
		&
		DDTI &
		77.327±10.665 &
		76.945±9.795 &
		73.529±14.75 &
		74.173±12.848 &
		73.441±16.719 &
		69.403±16.68 &
		68.014±14.24 &
		68.386±14.854 &
		68.444±18.15 &
		67.724±17.196 \\
		&
		&
		Benign &
		78.325±9.723 &
		77.301±11.218 &
		76.299±13.836 &
		74.566±16.745 &
		73.579±16.757 &
		72.466±15.702 &
		71.407±14.8 &
		72.065±15.868 &
		73.582±14.682 &
		70.922±15.718 \\
		&
		\multirow{3}{*}{Ablation:} &
		2 Sources &
		70.71±15.348 &
		69.882±14.456 &
		68.524±14.781 &
		69.324±14.773 &
		68.949±12.424 &
		65.375±8.094 &
		64.771±9.391 &
		62.777±10.024 &
		62.308±12.791 &
		52.052±12.541 \\
		&
		&
		DDTI &
		73.022±10.728 &
		72.038±9.067 &
		68.898±10.048 &
		71.357±17.074 &
		69.806±8.078 &
		70.308±8.916 &
		70.374±5.567 &
		67.772±9.338 &
		67.122±9.804 &
		67.106±6.28 \\
		&
		&
		Benign &
		65.024±14.173 &
		63.009±8.113 &
		63.35±7.701 &
		63.784±6.636 &
		62.726±11.198 &
		61.763±9.379 &
		61.49±8.445 &
		61.248±6.827 &
		62.374±8.675 &
		59.846±6.324 \\
		&
		\multicolumn{2}{c}{Only Target} &
		68.278±16.066 &
		64.945±12.027 &
		63.278±11.149 &
		64.662±8.122 &
		60.889±11.627 &
		62.717±14.265 &
		60.535±7.04 &
		59.895±6.049 &
		60.463±5.603 &
		57.607±6.562 \\
		&
		\multicolumn{2}{c}{Without Independence} &
		65.047±12.224 &
		66.136±14.06 &
		67.389±13.377 &
		67.799±12.699 &
		69.056±12.895 &
		68.633±13.648 &
		69.36±14.561 &
		69.687±16.365 &
		68.924±16.361 &
		70.297±16.871 \\ \hline
	\end{tabular}
}
\end{table*}
Overall, the four indicators of the BUSI(Malignant) experiment are overall higher than those of BUSI(Benign).
It can be seen that in the process of increasing the proportion of unlabeled data from $10\%$ to $90\%$, the effect of adversarial transfer learning using two source domains is significantly better than other experimental results.
It shows that the multi-source adversarial transfer learning applied to the BUSI( Malignant) dataset shows better performance.
The multi-source adversarial transfer learning strategy still performs the best.
The BUSI(Malignant) is better than the BUSI(Benign) experiment, the multi-source adversarial transfer learning metric decreases less as the proportion of labeled samples decreases. Because these two labeled source domains can provide more comprehensive deep features, high accuracy can be achieved without too many labels in the segmented part.
When using only BUSI(Benign) as the source domain, it can be seen that the accuracy is close but slightly lower than multi-source adversarial transfer learning.
It shows that although the source domain can provide most of the knowledge information of the target domain segmentation task, it does not yet contain all the knowledge required by the target domain;
When only DDTI is used as the source domain, and the unlabeled ratio is about $40\%\mbox{-}50\%$, the performance begins to drop significantly, and more target domain labels are needed to screen out features that are more beneficial to the target task through confrontation;
For deep learning that only uses the target domain, the performance drops significantly at about $40\%\mbox{-}50\%$ and $80\%\mbox{-}90\%$.
This is due to overfitting due to insufficient labeled training data.
The effect of adding an adversarial structure is also significantly better than that of multi-task learning without an adversarial structure (including one target domain and two source domains for a total of three tasks).
It can be considered that the common features of adversarial structure extraction are more conducive to the improvement of target domain tasks.
In addition, it can also be seen that the comparative experiments of adversarial transfer learning of two single source domains are weaker than multi-source adversarial transfer learning at all times. It can be considered that the information provided by multiple source domains is more than that of a single-source domain.
The overall effect of ablation experiments is not as good as that of adversarial transfer learning, and even some experimental performances are not as good as the performance of using only the target domain data, indicating that adding domain classifiers is beneficial to extract common features that improve the performance of target domain tasks.
\subsection{Parameter ablation experiment results and analysis}
We also refer to the work \cite{zhang2021transfer} to explore some key parameters. For the parameters $\alpha$ and $\lambda$ of the equation \eqref{losstotal}, the influence of different weight coefficients on the performance of the model is studied, and an ablation experiment is designed. We believe that the parameter weights $\alpha$ and $\lambda$ can be easily set according to specific needs in practical application scenarios. Specifically, we randomly prepare several sets of weights for testing on two target domain datasets.
\subsubsection{BUSI(Benign) parameter ablation experiment results and analysis}
\begin{figure}[h]
	\centerline{\includegraphics[width=\columnwidth]{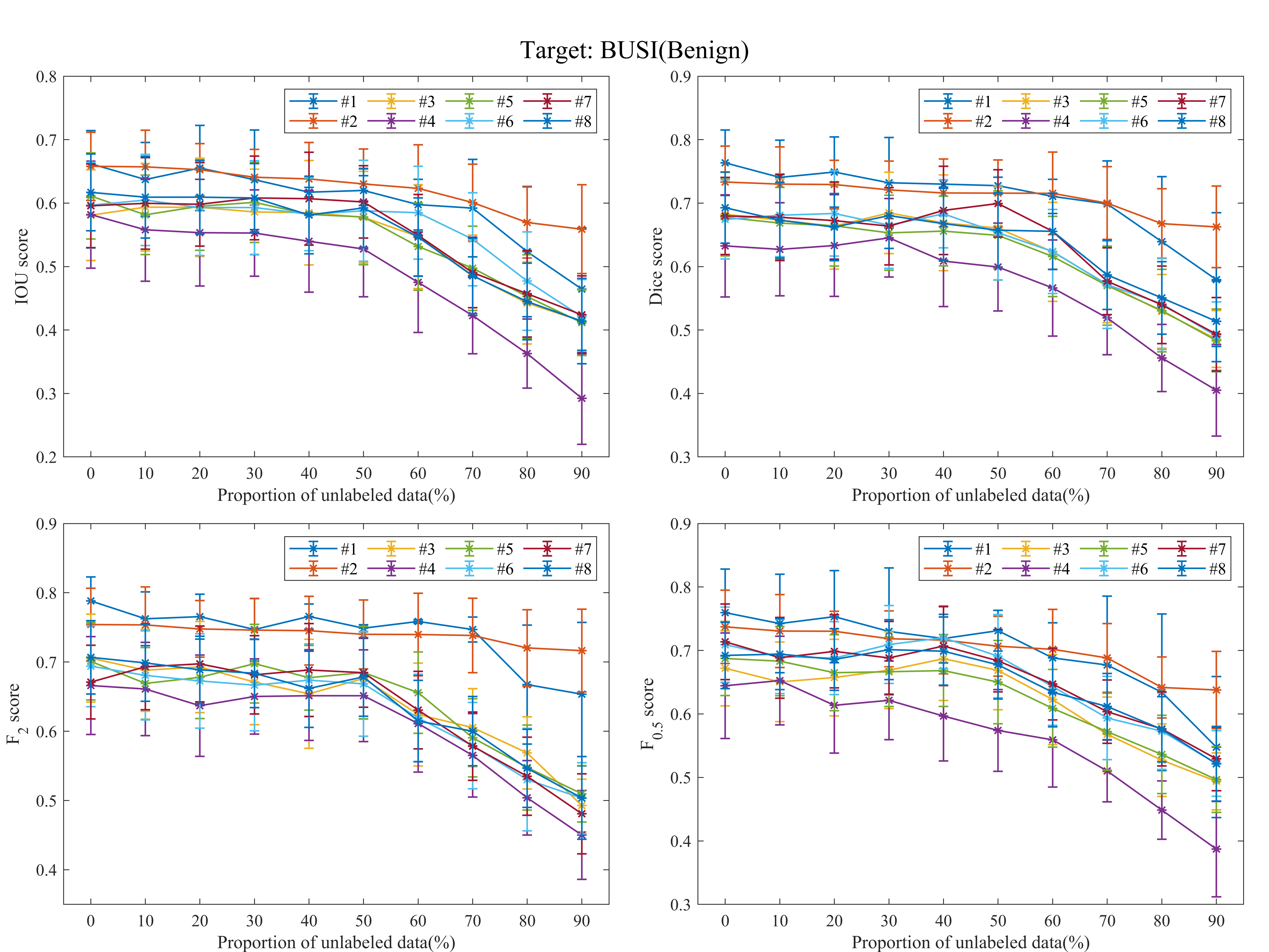}}
	\caption{Parametric ablation experiments using BUSI(Benign) as the target domain, and BUSI(Malignant) and CT2US as the two source domains. The curves of the four indicators of $IOU$, $Dice \ score$, $F_{2}$ and $F_{0.5}$ as a function of the proportion of unlabeled data are plotted. }
	\label{Benign_paramenters}
\end{figure}

We tested the performance metrics $IOU$, $Dice \ score$, $F_{2}$ and $F_{0.5}$ on the BUSI(Benign) dataset with different proportions of unlabeled data. The results are shown in the table \ref{Parameters_Benign_score_result}. Figure \ref{Benign_paramenters} plots the trend of the proportion of unlabeled data for all experiments in the table \ref{Parameters_Benign_score_result} from $0\%$ to $90\%$.
It can be seen that among the four indicators, \#1 and \#2 have the best performance. Specifically, \#1 has better performance when there are more labeled data and less unlabeled data. When the labeled data is less and the unlabeled data is more, the performance of \#2 is better. It can be observed that the performance of \#4 begins to drop significantly when the unlabeled data reaches $50\%$. We think this is because the parameter $\lambda$ used for confrontation is too small. The difference is that although \#6 performs better when there are fewer tags, its performance is generally not as good as \#1 and \#2. We believe this is due to the fact that the split predictors are underweighted. BUSI(Benign) parameter ablation experiments show that, in general, performance is better with different scales of labeled data when the weights are within a reasonable range of values close to each other. If the weight gap is too large, the error will increase accordingly when it is largely biased towards certain values.
\begin{table*}[thp]
	\tiny  
	\centering 
	\caption{BUSI(Benign) parameter ablation experiment results}   
	\label{Parameters_Benign_score_result}
	\resizebox{\textwidth}{!}
	{
		\begin{tabular}{clllllllllllll}
			\hline
			\multicolumn{4}{c}{Target: BUSI(Benign)} &
			\multicolumn{10}{c}{Proportion of unlabeled data} \\
			\multicolumn{1}{c}{Score(\%)} &
			Group &
			$\alpha$ &
			$\lambda$ &
			0\% &
			10\% &
			20\% &
			30\% &
			40\% &
			50\% &
			60\% &
			70\% &
			80\% &
			90\% \\ \hline
			\multirow{8}{*}{$IoU$} &
			\#1 &
			1 &
			1 &
			66.197±5.239 &
			63.725±5.832 &
			65.545±6.714 &
			63.68±7.845 &
			61.718±0.76 &
			61.986±2.343 &
			59.749±3.998 &
			59.216±7.682 &
			52.359±10.259 &
			46.45±9.653 \\
			&
			\#2 &
			2 &
			3 &
			65.79±5.383 &
			65.719±5.785 &
			65.267±4.098 &
			64.077±4.384 &
			63.821±5.749 &
			63.031±5.486 &
			62.331±6.865 &
			60.076±6.057 &
			56.94±5.592 &
			55.888±7.002 \\
			&
			\#3 &
			1 &
			4 &
			58.098±7.153 &
			59.359±6.911 &
			59.329±7.732 &
			58.595±6.752 &
			58.486±8.226 &
			57.773±7.217 &
			54.722±8.185 &
			48.599±6.34 &
			44.254±6.459 &
			41.404±4.954 \\
			&
			\#4 &
			3 &
			1 &
			58.174±8.429 &
			55.787±8.093 &
			55.336±8.398 &
			55.284±6.803 &
			53.974±8.009 &
			52.775±7.549 &
			47.51±7.907 &
			42.267±6.022 &
			36.283±5.456 &
			29.235±7.246 \\
			&
			\#5 &
			5 &
			3 &
			61.12±6.789 &
			58.161±6.255 &
			59.512±6.952 &
			60.107±6.314 &
			58.199±5.691 &
			57.827±7.537 &
			53.153±6.873 &
			49.734±6.649 &
			45.276±6.597 &
			41.141±5.162 \\
			&
			\#6 &
			1 &
			5 &
			59.66±6.784 &
			60.495±7.19 &
			59.288±7.553 &
			59.301±7.389 &
			58.242±5.76 &
			58.784±7.954 &
			58.488±7.308 &
			54.308±7.338 &
			47.704±7.772 &
			42.107±5.86 \\
			&
			\#7 &
			5 &
			2 &
			59.575±6.594 &
			59.954±7.215 &
			59.804±6.587 &
			60.812±6.596 &
			60.689±7.321 &
			60.19±5.704 &
			54.931±6.414 &
			49.046±5.546 &
			45.711±6.822 &
			42.394±6.153 \\
			&
			\#8 &
			2 &
			1 &
			61.683±6.043 &
			60.929±6.435 &
			60.926±5.808 &
			60.826±5.392 &
			58.116±6.112 &
			59.25±6.234 &
			54.65±6.171 &
			48.528±5.942 &
			44.489±6.032 &
			41.406±6.733 \\ \hdashline[1pt/1pt]
			\multirow{8}{*}{$Dice$} &
			\#1 &
			1 &
			1 &
			76.361±5.16 &
			74.044±5.86 &
			74.893±5.53 &
			73.212±7.127 &
			72.99±0.663 &
			72.743±1.347 &
			71.055±2.694 &
			69.868±6.797 &
			63.931±10.241 &
			57.94±10.535 \\
			&
			\#2 &
			2 &
			3 &
			73.337±5.639 &
			72.997±5.855 &
			72.917±3.824 &
			72.082±4.538 &
			71.597±5.361 &
			71.571±5.224 &
			71.539±6.517 &
			70.008±5.741 &
			66.772±5.494 &
			66.256±6.432 \\
			&
			\#3 &
			1 &
			4 &
			68.238±6.589 &
			67.317±6.434 &
			66.365±6.78 &
			68.467±6.41 &
			66.894±7.54 &
			66.066±6.44 &
			62.315±7.794 &
			57.189±6.039 &
			52.95±5.806 &
			48.581±4.482 \\
			&
			\#4 &
			3 &
			1 &
			63.265±8.044 &
			62.721±7.325 &
			63.313±8.025 &
			64.532±6.182 &
			60.859±7.172 &
			59.916±6.93 &
			56.619±7.568 &
			51.925±5.823 &
			45.584±5.299 &
			40.506±7.215 \\
			&
			\#5 &
			5 &
			3 &
			67.745±6.035 &
			66.85±5.64 &
			66.433±6.343 &
			65.321±5.916 &
			65.596±5.488 &
			64.91±7.029 &
			61.593±6.32 &
			57±6.237 &
			53.064±6.512 &
			48.353±4.953 \\
			&
			\#6 &
			1 &
			5 &
			67.439±6.248 &
			68.088±6.528 &
			68.362±6.706 &
			66.519±6.815 &
			68.318±5.372 &
			65.344±7.451 &
			62.378±6.63 &
			57.133±6.88 &
			54.123±7.217 &
			49.041±5.387 \\
			&
			\#7 &
			5 &
			2 &
			67.964±6.079 &
			67.748±6.754 &
			67.248±6.083 &
			66.363±6.096 &
			68.845±6.948 &
			69.946±5.314 &
			65.55±5.999 &
			57.699±5.261 &
			53.958±6.109 &
			49.334±5.794 \\
			&
			\#8 &
			2 &
			1 &
			69.283±5.613 &
			67.302±5.959 &
			66.221±5.292 &
			68.035±5.185 &
			66.766±5.754 &
			65.735±5.905 &
			65.558±6.014 &
			58.696±5.466 &
			55.068±5.719 &
			51.371±6.344 \\ \hdashline[1pt/1pt]
			\multirow{8}{*}{$F_2$} &
			\#1 &
			1 &
			1 &
			78.823±3.467 &
			76.246±3.881 &
			76.548±3.234 &
			74.679±4.515 &
			76.567±1.82 &
			74.873±0.507 &
			75.847±0.254 &
			74.691±1.802 &
			66.746±8.591 &
			65.353±10.359 \\
			&
			\#2 &
			2 &
			3 &
			75.41±5.238 &
			75.353±5.5 &
			74.786±4.081 &
			74.61±4.554 &
			74.515±4.96 &
			73.993±4.979 &
			73.967±5.964 &
			73.828±5.381 &
			72.023±5.522 &
			71.657±5.96 \\
			&
			\#3 &
			1 &
			4 &
			70.556±6.342 &
			68.808±5.91 &
			69.277±6.583 &
			67.094±6.135 &
			65.42±7.872 &
			67.625±5.87 &
			62.412±7.422 &
			60.525±5.635 &
			56.87±5.207 &
			49.282±3.799 \\
			&
			\#4 &
			3 &
			1 &
			66.612±7.068 &
			66.095±6.736 &
			63.697±7.317 &
			65.015±5.432 &
			65.12±6.437 &
			65.134±6.632 &
			61.101±6.993 &
			56.524±6.036 &
			50.398±5.381 &
			45.025±6.412 \\
			&
			\#5 &
			5 &
			3 &
			70.085±5.566 &
			66.91±5.187 &
			67.791±5.943 &
			69.754±5.679 &
			67.725±4.659 &
			68.474±6.671 &
			65.566±5.873 &
			59.058±5.643 &
			54.765±6.139 &
			50.97±4.08 \\
			&
			\#6 &
			1 &
			5 &
			69.401±5.836 &
			68.064±6.437 &
			67.274±6.817 &
			66.66±6.6 &
			67.366±5.25 &
			66.85±7.57 &
			61.927±6.306 &
			57.931±6.243 &
			53.029±7.402 &
			50.321±5.144 \\
			&
			\#7 &
			5 &
			2 &
			67.102±5.318 &
			69.311±6.201 &
			69.724±5.469 &
			68.155±5.653 &
			68.848±6.698 &
			68.448±4.977 &
			63.042±5.599 &
			57.845±4.932 &
			53.507±5.649 &
			48.086±5.774 \\
			&
			\#8 &
			2 &
			1 &
			70.669±5.295 &
			69.88±5.562 &
			68.868±4.864 &
			68.363±4.941 &
			66.183±5.609 &
			67.835±5.681 &
			61.475±5.878 &
			60.018±4.996 &
			54.652±5.635 &
			50.373±5.977 \\ \hdashline[1pt/1pt]
			\multirow{8}{*}{$F_{0.5}$} &
			\#1 &
			1 &
			1 &
			75.972±6.84 &
			74.245±7.76 &
			75.318±7.253 &
			73.013±9.98 &
			71.873±3.821 &
			73.12±3.217 &
			68.812±5.543 &
			67.723±10.848 &
			63.42±12.336 &
			54.783±11.118 \\
			&
			\#2 &
			2 &
			3 &
			73.721±5.784 &
			73.063±5.755 &
			73.024±3.152 &
			71.843±4.397 &
			71.648±5.333 &
			70.674±4.726 &
			70.18±6.3 &
			68.838±5.409 &
			64.186±4.807 &
			63.783±6.078 \\
			&
			\#3 &
			1 &
			4 &
			67.183±5.909 &
			65.056±6.275 &
			65.729±6.028 &
			66.88±6.043 &
			68.702±6.585 &
			66.862±6.192 &
			62.258±7.106 &
			56.788±5.874 &
			52.711±5.713 &
			49.377±4.499 \\
			&
			\#4 &
			3 &
			1 &
			64.449±8.312 &
			65.263±6.98 &
			61.357±7.515 &
			62.145±6.209 &
			59.672±7.07 &
			57.401±6.445 &
			55.915±7.43 &
			51.019±4.856 &
			44.857±4.594 &
			38.71±7.51 \\
			&
			\#5 &
			5 &
			3 &
			68.73±5.841 &
			68.306±5.435 &
			66.505±5.989 &
			66.636±5.451 &
			66.841±5.676 &
			65.014±6.566 &
			60.89±6.12 &
			57.211±6.056 &
			53.631±6.171 &
			49.671±5.181 \\
			&
			\#6 &
			1 &
			5 &
			70.906±5.924 &
			69.043±5.838 &
			68.792±5.742 &
			70.937±6.144 &
			72.009±4.862 &
			69.108±6.56 &
			64.32±6.104 &
			59.373±6.561 &
			57.291±6.039 &
			52.194±5.168 \\
			&
			\#7 &
			5 &
			2 &
			71.361±5.956 &
			68.851±6.359 &
			69.863±5.753 &
			68.842±5.719 &
			70.698±6.238 &
			68.377±4.915 &
			64.716±5.659 &
			60.365±4.999 &
			57.634±5.828 &
			52.919±5.019 \\
			&
			\#8 &
			2 &
			1 &
			69.206±5.21 &
			69.441±5.593 &
			68.571±4.863 &
			70.121±4.713 &
			69.889±5.37 &
			67.734±5.382 &
			63.414±5.433 &
			61.169±5.214 &
			57.58±5.136 &
			52.147±5.885 \\ \hline
		\end{tabular}
	}
\end{table*}
\subsubsection{BUSI(Malignant) parameter ablation experiment results and analysis}
\begin{figure}[h]
	\centerline{\includegraphics[width=\columnwidth]{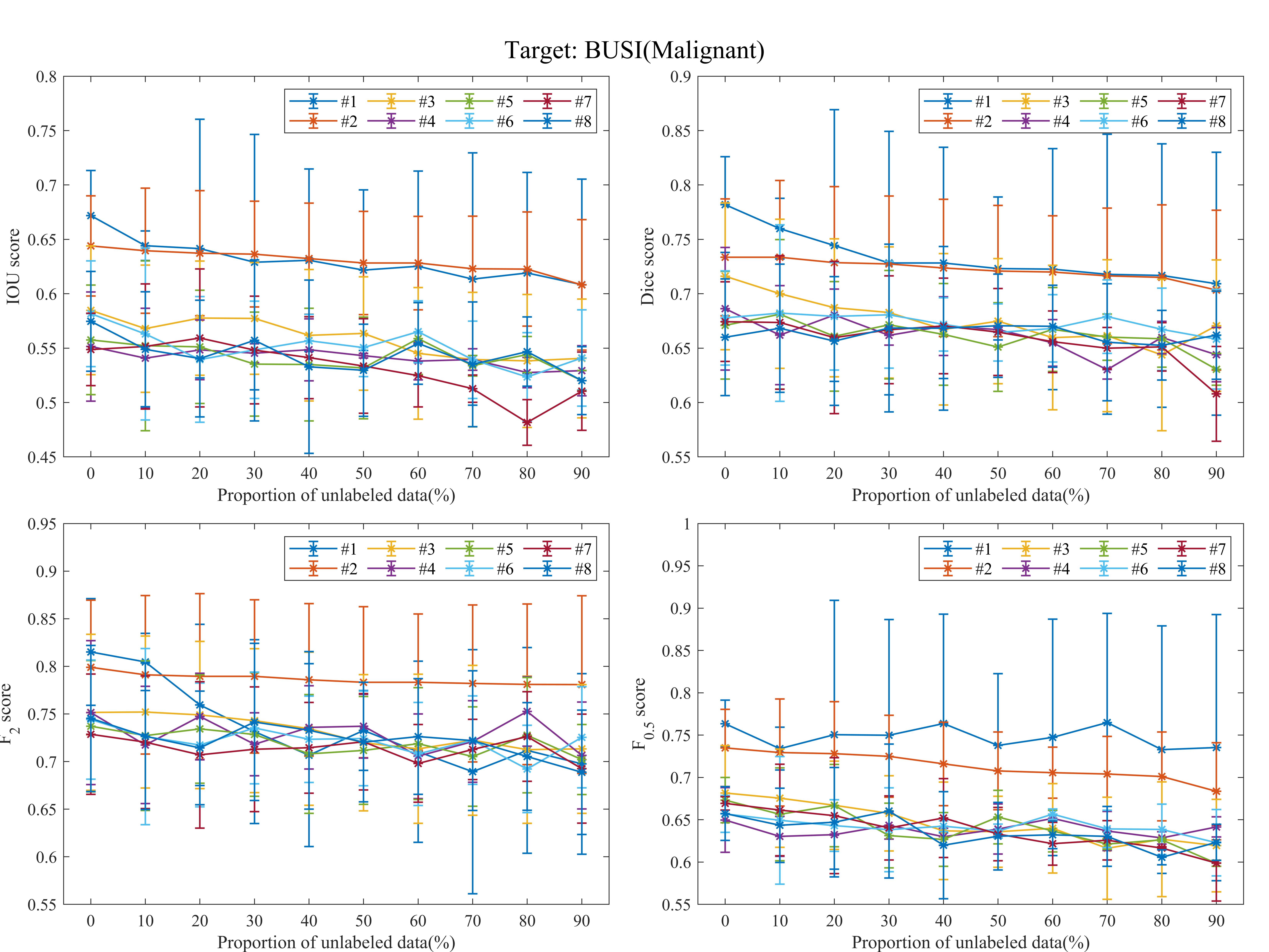}}
	\caption{Parametric ablation experiments using BUSI(Malignant) as the target domain, and BUSI(Benign) and DDTI as the two source domains. The curves of the four indicators of $IOU$, $Dice \ score$, $F_{2}$ and $F_{0.5}$ as a function of the proportion of unlabeled data are plotted. }
	\label{Malignant_paramenters}
\end{figure}

Similar to the parameter ablation experiment of BUSI (Benign), the figure \ref{Malignant_paramenters} draws the trend of the test results in the table \ref{Parameters_Malignant_score_result}. Like BUSI(Benign), \#1 and \#2 perform best. It is worth noting that \#2 has very limited performance degradation as the number of labels decreases. We believe that under the condition that the weight of the segmentation predictor is not too small, the coefficient of the corresponding item of the domain discriminator in the loss function can be appropriately increased within a certain range to achieve better performance with fewer labels.
\begin{table}[thp]
	\tiny  
	\centering 
	\caption{BUSI(Malignant) parameter ablation experiment results}   
	\label{Parameters_Malignant_score_result}
	\resizebox{\textwidth}{!}
	{
		\begin{tabular}{clllllllllllll}
			\hline
			\multicolumn{4}{c}{Target: BUSI(Malignant)} &
			\multicolumn{10}{c}{Proportion of unlabeled data} \\
			\multicolumn{1}{c}{Score(\%)} &
			Group &
			$\alpha$ &
			$\lambda$ &
			0\% &
			10\% &
			20\% &
			30\% &
			40\% &
			50\% &
			60\% &
			70\% &
			80\% &
			90\% \\ \hline
			\multirow{8}{*}{$IoU$} &
			\#1 &
			1 &
			1 &
			67.177±4.149 &
			64.422±1.358 &
			64.154±11.891 &
			62.906±11.746 &
			63.08±8.387 &
			62.168±7.378 &
			62.521±8.765 &
			61.352±11.598 &
			61.907±9.248 &
			60.841±9.704 \\
			&
			\#2 &
			2 &
			3 &
			64.394±4.609 &
			63.955±5.757 &
			63.727±5.748 &
			63.64±4.861 &
			63.227±5.106 &
			62.83±4.748 &
			62.826±4.297 &
			62.306±4.827 &
			62.27±5.243 &
			60.815±5.999 \\
			&
			\#3 &
			1 &
			4 &
			58.478±5.919 &
			56.787±5.854 &
			57.757±5.245 &
			57.735±5.132 &
			56.183±6.038 &
			56.346±5.207 &
			54.524±6.054 &
			53.952±6.183 &
			53.824±6.116 &
			54.053±5.456 \\
			&
			\#4 &
			3 &
			1 &
			55.148±5.021 &
			54.071±4.598 &
			54.839±2.742 &
			54.531±0.876 &
			54.847±2.847 &
			54.307±0.21 &
			53.814±1.731 &
			53.953±0.989 &
			52.749±1.395 &
			52.921±2.311 \\
			&
			\#5 &
			5 &
			3 &
			55.761±5.037 &
			55.217±7.807 &
			55.117±5.204 &
			53.533±4.776 &
			53.486±5.174 &
			53.177±4.677 &
			55.874±3.298 &
			53.38±0.938 &
			54.414±1.662 &
			52.014±0.934 \\
			&
			\#6 &
			1 &
			5 &
			58.161±4.854 &
			56.297±7.908 &
			53.956±5.778 &
			54.839±4.474 &
			55.674±2.428 &
			55.04±2.657 &
			56.503±2.847 &
			53.927±3.563 &
			52.353±4.074 &
			54.092±4.433 \\
			&
			\#7 &
			5 &
			2 &
			54.879±3.33 &
			55.16±5.752 &
			55.937±6.335 &
			54.822±4.953 &
			54.125±3.768 &
			53.358±4.348 &
			52.478±2.882 &
			51.269±1.255 &
			48.16±2.097 &
			51.042±3.603 \\
			&
			\#8 &
			2 &
			1 &
			57.467±4.592 &
			54.891±5.286 &
			54.04±5.362 &
			55.709±7.405 &
			53.29±7.958 &
			52.972±4.233 &
			55.439±3.757 &
			53.516±5.738 &
			54.68±3.187 &
			52.009±3.126 \\ \hdashline[1pt/1pt]
			\multirow{8}{*}{$Dice$} &
			\#1 &
			1 &
			1 &
			78.194±4.405 &
			75.988±2.793 &
			74.431±12.49 &
			72.827±12.112 &
			72.83±10.631 &
			72.317±6.573 &
			72.262±11.084 &
			71.795±12.875 &
			71.676±12.112 &
			70.921±12.085 \\
			&
			\#2 &
			2 &
			3 &
			73.355±5.373 &
			73.354±7.054 &
			72.859±6.995 &
			72.735±6.255 &
			72.377±6.316 &
			72.087±6.037 &
			72.003±5.164 &
			71.64±6.24 &
			71.511±6.668 &
			70.334±7.351 \\
			&
			\#3 &
			1 &
			4 &
			71.607±6.751 &
			70±6.852 &
			68.708±6.335 &
			68.274±6.041 &
			66.727±6.961 &
			67.483±5.746 &
			65.968±6.642 &
			66.149±6.987 &
			64.372±6.957 &
			67.062±6.046 \\
			&
			\#4 &
			3 &
			1 &
			68.622±5.642 &
			66.192±4.55 &
			68.044±2.376 &
			66.17±0.888 &
			67.024±2.699 &
			66.711±0.203 &
			65.483±2.145 &
			63.019±0.862 &
			65.981±1.49 &
			64.4±2.5 \\
			&
			\#5 &
			5 &
			3 &
			67.099±4.952 &
			68.113±6.875 &
			66.079±5.045 &
			67.154±4.982 &
			66.262±4.669 &
			65.097±4.079 &
			66.721±3.855 &
			66.014±2.126 &
			65.858±2.605 &
			63.045±1.465 \\
			&
			\#6 &
			1 &
			5 &
			67.776±4.336 &
			68.226±8.113 &
			67.919±4.937 &
			68.072±4.913 &
			67.176±2.452 &
			66.443±2.632 &
			66.818±3.095 &
			67.9±3.384 &
			66.731±3.785 &
			65.824±4.59 \\
			&
			\#7 &
			5 &
			2 &
			67.436±3.661 &
			67.36±6.141 &
			65.971±6.995 &
			66.689±4.951 &
			67.042±4.391 &
			66.477±3.995 &
			65.585±2.832 &
			64.999±1.907 &
			65.118±2.22 &
			60.798±4.355 \\
			&
			\#8 &
			2 &
			1 &
			65.996±5.358 &
			66.836±5.892 &
			65.655±5.917 &
			66.847±7.715 &
			66.822±7.522 &
			67.059±4.745 &
			67.009±3.764 &
			65.546±5.381 &
			65.265±3.203 &
			66.187±4.053 \\ \hdashline[1pt/1pt]
			\multirow{8}{*}{$F_2$} &
			\#1 &
			1 &
			1 &
			81.505±5.605 &
			80.462±2.995 &
			75.952±8.46 &
			73.144±9.648 &
			70.675±9.604 &
			73.262±4.19 &
			71.033±9.51 &
			68.938±12.816 &
			71.17±10.811 &
			69.753±9.483 \\
			&
			\#2 &
			2 &
			3 &
			79.89±7.065 &
			79.114±8.326 &
			78.946±8.699 &
			78.944±8.057 &
			78.579±8.009 &
			78.334±7.927 &
			78.32±7.178 &
			78.208±8.246 &
			78.112±8.431 &
			78.078±9.328 \\
			&
			\#3 &
			1 &
			4 &
			75.169±8.188 &
			75.203±7.979 &
			74.892±7.73 &
			74.29±7.554 &
			73.449±8.045 &
			71.97±7.153 &
			71.336±7.83 &
			72.23±7.867 &
			71.245±7.731 &
			71.33±6.761 \\
			&
			\#4 &
			3 &
			1 &
			75.133±7.556 &
			71.745±6.154 &
			74.733±4.545 &
			71.815±3.297 &
			73.593±4.369 &
			73.714±3.347 &
			70.557±4.45 &
			72.108±4.275 &
			75.265±3.654 &
			70.643±5.611 \\
			&
			\#5 &
			5 &
			3 &
			73.713±6.899 &
			72.72±7.86 &
			73.42±5.691 &
			72.849±6.49 &
			70.803±6.237 &
			71.171±5.671 &
			71.883±5.871 &
			70.531±5.22 &
			72.769±6.056 &
			70.212±3.68 \\
			&
			\#6 &
			1 &
			5 &
			74.408±6.264 &
			72.618±9.255 &
			71.741±6.496 &
			73.524±5.868 &
			72.34±4.527 &
			72.427±4.973 &
			70.799±5.419 &
			72.249±4.652 &
			69.225±4.585 &
			72.544±5.31 \\
			&
			\#7 &
			5 &
			2 &
			72.871±6.315 &
			72.035±6.961 &
			70.7±7.694 &
			71.295±6.556 &
			71.443±4.765 &
			72.077±5.059 &
			69.81±4.086 &
			71.269±3.17 &
			72.624±4.706 &
			69.256±5.724 \\
			&
			\#8 &
			2 &
			1 &
			74.532±7.671 &
			72.655±7.687 &
			71.434±5.969 &
			74.159±8.257 &
			73.279±8.279 &
			72.044±6.275 &
			72.624±6.078 &
			72.198±7.339 &
			70.528±5.654 &
			68.876±6.524 \\ \hdashline[1pt/1pt]
			\multirow{8}{*}{$F_{0.5}$} &
			\#1 &
			1 &
			1 &
			76.354±2.787 &
			73.403±2.529 &
			75.043±15.874 &
			74.982±13.667 &
			76.376±12.921 &
			73.777±8.488 &
			74.74±13.963 &
			76.469±12.901 &
			73.277±14.622 &
			73.53±15.722 \\
			&
			\#2 &
			2 &
			3 &
			73.492±4.562 &
			72.935±6.324 &
			72.807±6.154 &
			72.496±4.841 &
			71.594±4.928 &
			70.77±4.591 &
			70.563±3.01 &
			70.409±4.435 &
			70.12±5.252 &
			68.363±5.733 \\
			&
			\#3 &
			1 &
			4 &
			68.166±5.604 &
			67.546±5.805 &
			66.7±5.218 &
			65.754±4.435 &
			63.691±5.764 &
			63.584±4.201 &
			63.981±5.286 &
			61.622±6.03 &
			62.693±6.783 &
			61.954±5.463 \\
			&
			\#4 &
			3 &
			1 &
			64.973±3.814 &
			63.032±2.356 &
			63.24±0.067 &
			64.343±1.623 &
			62.983±1.015 &
			63.904±2.953 &
			65.164±0.352 &
			63.66±2.314 &
			62.848±0.726 &
			64.152±1.188 \\
			&
			\#5 &
			5 &
			3 &
			67.319±2.665 &
			65.633±5.487 &
			66.679±4.869 &
			63.135±3.826 &
			62.667±3.169 &
			65.321±3.166 &
			63.663±2.461 &
			62.131±0.639 &
			62.623±0.127 &
			59.87±0.365 \\
			&
			\#6 &
			1 &
			5 &
			65.688±2.171 &
			64.932±7.535 &
			64.308±3.052 &
			63.817±4.965 &
			64.24±0.629 &
			63.656±0.835 &
			65.666±0.618 &
			63.925±2.195 &
			63.859±2.985 &
			62.291±3.919 \\
			&
			\#7 &
			5 &
			2 &
			66.932±0.789 &
			66.15±5.402 &
			65.491±6.844 &
			64.032±3.78 &
			65.201±4.667 &
			63.301±3.151 &
			62.165±2.522 &
			62.577±2.32 &
			61.663±0.138 &
			59.876±4.487 \\
			&
			\#8 &
			2 &
			1 &
			65.728±3.169 &
			64.35±4.387 &
			64.705±6.463 &
			66.02±7.911 &
			61.986±6.327 &
			63.051±3.992 &
			63.213±1.625 &
			63.037±3.536 &
			60.555±0.882 &
			62.324±2.13 \\ \hline
		\end{tabular}
	}
\end{table}
\par
In summary, multi-source adversarial transfer learning can increase the generalization ability of the model through the strategy of transfer learning;
Through the method of adversarial transfer learning, some common features that can improve the performance of the target domain can be extracted from some source domains that have a certain appearance gap with the target domain;
If the key features provided by a source domain that are beneficial to the target domain task are not sufficient, the missing key features can be filled by multiple source domains with different similarities.
Through two comparative experiments, it can be seen that the proposed multi-source adversarial transfer learning method has better potential when the similarity between the source domain ultrasound image and the target domain ultrasound image is insufficient, and only some local features are similar. The ablation experiments prove that in the face of such a scene with insufficient global similarity but local similarity, the proposed multi-source adversarial transfer learning can extract more source domains to assist target domain learning to obtain better performance knowledge. Can handle this scenario better. Through parameter ablation experiments, we found that the weight of the segmentation predictor can be appropriately increased when there is more labeled data, and the weight of the domain classifier should be appropriately increased when the labeled data is insufficient. The weights of both are very important, and a set of relatively close parameters $\alpha$ and $\lambda$ can be selected, because unbalanced parameters will lead to a decrease in the performance of the proposed method.
\section{CONCLUSION}
We propose a novel multi-source adversarial transfer learning network for the task of source-domain ultrasound image segmentation with limited similarity to the target domain.
Specifically, through the strategy of adversarial transfer learning, general deep features beneficial to target domain tasks are adaptively extracted from different source domains, respectively, and the unlabeled target domain data is maximized.
By leveraging knowledge from multiple source domains to supplement key features lacking in a single source domain and target domain due to lack of similarity, a multi-source transfer learning strategy is used to optimize task performance as much as possible.
The segmentation results on three organ datasets confirm the good transfer learning ability of multi-source adversarial transfer learning when the proportion of labeled data is insufficient and when the domain similarity is limited.
\section*{Acknowledgements}

This work was supported in part by the National Natural Science Foundation of China (61973067).

\bibliography{wpref}

\end{document}